\documentclass[twocolumn]{aastex631}

\usepackage{apjfonts}

\DeclareSymbolFont{cmletters}{OML}{cmm}{m}{it}
\DeclareMathSymbol{v}{\mathalpha}{cmletters}{"76}

\newcommand{\hammer}{H-AMR}

\definecolor{nick}{HTML}{006400}

\shorttitle{Precessing Accretion Disks}
\shortauthors{Liska et al.}
\graphicspath{{./}{figures/}}

\begin{document}

\title{Radiation Transport Two-Temperature GRMHD Simulations of Warped Accretion Disks}

\correspondingauthor{Matthew Liska}
\email{matthew.liska@cfa.harvard.edu}
\author{M.T.P. Liska}
\affiliation{Institute for Theory and Computation, Harvard University, 60 Garden Street, Cambridge, MA 02138, USA}
\author[0000-0002-5375-8232]{N. Kaaz}
\affiliation{Center for Interdisciplinary Exploration \& Research in Astrophysics (CIERA), Physics \& Astronomy, Northwestern University, Evanston, IL 60202, USA}
\author{G. Musoke}
\affiliation{Anton Pannekoek Institute for Astronomy, University of Amsterdam, Science Park 904, 1098 XH Amsterdam, The Netherlands}
\author[0000-0002-9182-2047]{A. Tchekhovskoy}
\affiliation{Center for Interdisciplinary Exploration \& Research in Astrophysics (CIERA), Physics \& Astronomy, Northwestern University, Evanston, IL 60202, USA}
\author[0000-0002-4584-2557]{O. Porth}
\affiliation{Anton Pannekoek Institute for Astronomy, University of Amsterdam, Science Park 904, 1098 XH Amsterdam, The Netherlands}



\begin{abstract}
In many black hole systems, the accretion disk is expected to be misaligned with respect to the black hole spin axis. If the scale height of the disk is much smaller than the misalignment angle, the spin of the black hole can tear the disk into multiple, independently precessing `sub-disks'. This is most likely to happen during outbursts in black hole X-Ray binaries (BHXRBs) and in active galactic nuclei (AGN) accreting above a few percent of the Eddington limit, because the disk becomes razor-thin. Disk tearing has the potential to explain variability phenomena including quasi-periodic oscillations (QPOs) in BHXRBs and changing-look phenomena in AGN. Here, we present the first radiative two-temperature GRMHD simulation of a strongly tilted ($65^{\circ}$) accretion disk around a $M_{BH}=10M_{\odot}$ black hole, which tears and precesses. This leads to luminosity swings  between a few percent and $50 \%$ of the Eddington limit on sub-viscous timescales. Surprisingly, even where the disk is radiation pressure dominated, the accretion disk is thermally stable over $t \gtrsim 21,000 r_g/c$. This suggests warps play an important role in stabilizing the disk against thermal collapse. The disk forms two nozzle shocks perpendicular to the line of nodes where the scale height of the disk decreases $10$-fold and the electron temperature reaches $T_e \sim 10^8-10^9 K$. In addition, optically thin gas crossing the tear between the inner and outer disk gets heated to $T_e \sim 10^8 K$. This suggests that warped disks may emit a Comptonized spectrum that deviates substantially from idealized models.
\end{abstract}


\section{Introduction}
\label{sec:Intro}
Accretion disks in black hole X-ray binaries (BHXRBs) and active galactic nuclei (AGN) are typically described by axisymmetric, time-independent models. A key factor that determines the most suitable model is the accretion rate with respect to the Eddington limit. At the Eddington limit, radiation pressure rivals gravity and plays a dynamically important role. The most well-known models are the geometrically thin standard accretion disk model \citep{ss73}, which describes sources accreting at an appreciable fraction of their Eddington limit ($L \gtrsim 0.01-0.1\,L_{\rm Edd}$), and the geometrically thick advection dominated accretion flow model \citep{Narayan1994, Narayan2003}, which describes highly sub-Eddington sources ($L \lesssim 0.01 L_{\rm Edd}$). Slim disk models, in which the advection of trapped photons dominates, work well near and above the Eddington limit (e.g. \citealt{Abramowicz1978}). While these models can provide a reasonable description of the multi-wavelength emission, they -- by construction -- do not address the variability of their light curves and spectra. Such variability can encode important information about the structure of the accretion disk and corona (e.g. \citealt{Kara2019,UttleyCackett2014}).

For example, recent observational evidence indicates that accretion disks in changing-look AGN can undergo drastic changes in luminosity and spectral shape over the course of less than a year (e.g. \citealt{Maclead2016, Yang2018}) or during quasi-periodic eruptions (QPEs) over the course of hours (e.g. \citealt{Qian2018}). In these AGN, the luminosity can change by up to two orders of magnitude while the spectrum significantly hardens or softens. This suggest a dramatic change in the accretion flow on timescales which are too short to be explained by the viscous timescales inferred from the Maxwell stresses created by magnetorotational instability (MRI, \citealt{Balbus1991, Balbus1998}) induced turbulence (e.g. \citealt{Lawrence2018, Dexter2019}). Various physical mechanisms were proposed to reconcile theory with observations, including instabilities in the accretion disk \citep{Sniegowska2020}, magnetically elevated accretion disks \citep{Begelman2007}, reprocessing of point-source radiation by the outer accretion disk (e.g. \citealt{Clavel1992}), disk tearing \citep{Raj2021_AGN} and an orbiting compact object (e.g. \citealt{King2020, Arcodia2021}).

Similar to their changing-look AGN counterparts, the power spectra of BHXRBs display a wide range of mysterious features ranging from broad-spectrum variability to narrow spectral peaks known as quasi-periodic oscillations (QPOs). Such features can encode unique information about the structure and dynamics of the accretion disk in addition to the spin and mass of the black hole. QPOs are typically divided into low- and high-frequency QPOs, which can sometimes be observed together (e.g. \citealt{IngramMotta2020}). Various physical mechanisms have been proposed to explain QPOs. These include geometric effects such as precession of a misaligned disk around the black hole spin axis (e.g. \citealt{Stella1998, Ingram2009}) and intrinsic effects such as parametric resonances (e.g. \citealt{Kluzniak2002, Rezzolla2003, Abramowicz2003}) and disco-seismic modes (e.g. \citealt{Kato2004, Dewberry2019, Dewberry2020}). All of these models make simplifying assumptions about the underlying physics. Most significantly, they typically do not include magnetized turbulence (though see \citealt{Dewberry2020, Wagoner2021}), which can potentially dampen oscillatory modes. Thus, general relativistic magnetohydrodynamic (GRMHD) simulations, which can simulate accretion from first-principles, are very attractive for addressing the origin of QPOs.

Interestingly, numerous GRMHD simulations have demonstrated that a misaligned disk can precess (e.g. \citealt{Fragile2005, Fragile2007, Teixeira2014, Liska2018A, Liska2019A, White2019A}). However, this only happens when the disk is unrealistically small and, in fact, \citet{Liska2018A} demonstrated that the precession period of more realistically sized disks ($r \gtrsim few \times 100 r_g$) becomes too long to explain any observed QPOs. This problem can be solved if a smaller precessing disk tears of from a larger non precessing disk due to the differential frame dragging \citep{LT1918} of space-time by a spinning black hole. This process, called disk tearing, was observed in various SPH simulations (e.g. \citealt{Nixon2012B, Nixon2012counter, Nealon2015, Raj2021}) and recently confirmed by GRMHD simulations \citep{Liska2019b, Liska2020}. Disk tearing, interstingly, also leads to rapid burst of accretion possibly explaining large luminosity and spectral swings observed in accreting black holes (e.g. \citealt{Nixon2014, Raj2021_AGN}).

\citet{Musoke2022} recently addressed the origin of both low and high-frequency QPOs using GRMHD simulations where a $65^{\circ}$ misaligned accretion disk tore apart. Based on these simulations, \citet{Musoke2022} argued that low-frequency and high-frequency QPOs can be explained by disk tearing. In their model, a disk tears off at a near-constant radius ($r \sim 10^1 r_g$) and precesses for $\lesssim 10$ periods before falling into the black hole. While precessing at $\nu=2.5$~Hz (for $M_{BH}=10M_{\odot}$), the disk emits a periodically modulated light curve.  \citet{Musoke2022} also found a prominent radial epicyclic oscillations at $\nu \sim 56$~Hz in the inner disk, which they argue can potentially explain some high-frequency QPOs.  The geometric origin of low-frequency QPOs is supported by recent observations, which suggest that that iron line centroid frequency moves from blue-to-redshifted with the same phase as the low-frequency QPO \citep{Ingram2016_method, Ingram2016_QPO}. On the other hand, observations have been able to constrain neither an intrinsic nor geometric origin of the much rarer high-frequency QPOs (e.g. \citealt{IngramMotta2020}).

Thus, the results of \citet{Musoke2022} suggest that disk tearing is a very promising mechanism to explain the multi-wavelength variability in many accreting black holes. However, all GRMHD simulations that found tearing \citep{Liska2019b, Liska2020} have relied on a cooling function \citep{Noble2009} to artificially set the temperature and scale height of the disk. In reality, the temperature of the disk is determined by complex physics involving viscous heating, radiative cooling and the advection of energy. Such physics can modify the radial and vertical structure of the disk, such as its scale height and vertical temperature profile. This makes it extremely challenging to benchmark such GRMHD models against multi-wavelength observations because they might not capture the important physics that sets the disk thermodynamic state. For example, in radiation pressure supported disks imbalance between the local dissipation rate and cooling rate can lead to thermal and viscous instabilities that are not captured by a cooling function (e.g. \citealt{Lightman1974, Shakura1976, Sadowski2016_mag, Jiang2019, Liska2022}). In addition, when the inner disk precesses, it will beam its radiation field periodically towards the outer disk. This radiation field will scatter off the outer disk, and the radiation back-reaction force can cause additional warping and dissipation (e.g. \citealt{Pringle1997, Wijers1999}).

Here, we present the first radiation-transport two-temperature GRMHD simulation of an accretion disk tilted by $65^{\circ}$ relative to a rapidly spinning $M_{BH}=10M_{\odot}$ black hole of dimensionless spin of $a = 0.9375$ accreting at $35 \%$ the Eddington luminosity. We show that the disk undergoes tearing. The accretion disk does not contain any large scale vertical magnetic flux to begin with and does not launch a jet: this makes the simulation applicable to the soft(-intermediate) states of BHXRBs (see \citealt{FenderBelloni2004}). In Sections~\ref{sec:Numerics} and \ref{sec:Physical_Setup} we describe the numerical setup and initial conditions, before  presenting and discussing our results in Sections~\ref{sec:Results} and \ref{sec:discussion}, and concluding in Section~\ref{sec:conclusion}.

\section{Numerical Setup}
\label{sec:Numerics}
In this work we utilize the radiative version of our GPU accelerated GRMHD code \hammer{} \citep{Liska2018A, Liska2020}. H-AMR evolves the ideal GRMHD equations and radiation transfer equations in addition to the electron and ion entropies. We use a (modified) spherical grid in Kerr-Shild foliation with coordinates $r$, $\theta$ and $\phi$. Spatial reconstruction of primitive variables is performed using a 1-dimensional PPM method, which guarantees second order convergence in 3 dimensions. Magnetic fields are evolved on a staggered grid as described in \citet{Gardiner2005,White2016}. Inversion of conserved to primitive variables is performed using a 2-dimensional Newton-Raphson routine \citep{Noble2006} or Aitken Acceleration scheme \citep{Newman2014} for the energy equation and a 1-dimensional Newton-Raphson routine for the entropy equation. The energy based inversion is attempted first, and, if it fails, the entropy based inversion is used as a backup. This dual energy formulation is now standard in many GRMHD codes \citep{Porth2019}.  

The radiative transfer equations are closed with a two-moment M1-closure \citep{Levermore1984} whose specific implementation is described in \citet{Sadowski2013, Mckinney2013}. We additionally evolve (see e.g. \citealt{Noble2009, Ressler2015}) the electron and ion entropy tracers ($\kappa_{e,i}=p_{e,i}/\rho^{\gamma_{e,i}}$) and include Coulomb collisions \citep{Stepney1983}. We use adiabatic indices $\gamma_e=\gamma_i=5/3$ for the electrons and ions (and $\gamma=5/3$ for the gas). This is appropriate for the ion and electron temperatures in the accretion disk. For a typical `peak' temperature of $T_e \sim 10^9K$ and $T_i \sim 10^{10} K$ we find that $\gamma_e \sim 1.57$, $\gamma_{i} \sim 1.66$, $\gamma_g \sim 1.66$. We use a reconnection heating model \citep{Rowan2017} to distribute the dissipative heating between ions and electrons. This leads to roughly $20-40\%$ of the dissipation going into the electrons with the remainder heating the ions. 

We include Planck-averaged bound-free, free-free, and cyclo-Synchrotron absorption ($\kappa_{abs}$), emission ($\kappa_{em}$) and electron scattering ($\kappa_{es}$) opacities as given in \citet{Mckinney2017} for solar abundances $X=0.7$, $Y=0.28$, and $Z=0.02$. In addition, we account for Comptonization through a local blackbody approximation \citep{Sadowski2015}. This works reasonably well when emission, absorption and scattering are localized, but becomes inaccurate when the radiation field is anisotropic and/or the energy spectrum deviates from a blackbody. Namely, the radiation temperature used to calculate the absorption opacity is approximated as a blackbody with $T_{r}=(\hat{E}_{rad}/a)^{0.25}$ with $\hat{E}_{rad}$ the fluid frame radiation energy density and $a$ the radiation density constant. If the radiation deviates from a blackbody this approximation can severely under or overestimate the radiation temperature. We assume that only electrons can absorb and emit radiation and set the ion opacity to $\kappa_i=0$. This is a reasonable approximation in the accretion disk where bound-free processes, which involve energy exchange between ions/electrons/radiation, only become important when the density is high enough for Coulomb collisions to equilibrate the ion and electron temperature. The limitations of the M1 closure relevant to our work are further addressed in Section \ref{sec:discussion}. 

To resolve the accretion disk in our radiative GRMHD models we use a base grid resolution of $N_r \times N_{\theta} \times N_{\phi} = 840 \times 432 \times 288$. We then use beyond $r \gtrsim 4 r_g$ up to 3 layers of adaptive mesh refinement (AMR) to progressively increase this resolution in the disk from the base resolution at $r \sim 4 r_g$ to $N_r \times N_{\theta} \times N_{\phi} = 6720 \times 2304 \times 4096$ at $r \sim 10 r_g$. (In addition, we use static mesh refinement \citealt{Liska2020} to reduce the $\phi$-resolution to $N_{\phi}=16$ at the polar axis.) This is exactly half of the resolution used in \citet{Liska2020} and \citet{Musoke2022}. We place the radial boundaries at $R_{in}=1.13 r_g\approx 0.84r_{\rm H}$ and $R_{out}=10^5 r_g$, which ensures that both boundaries are causally disconnected from the flow; here $r_{\rm H} = 1+(1-a^2)^{1/2}r_g$ is the event horizon radius. The local adaptive time-stepping routine in \hammer{} increases the timestep of each mesh block independent of the refinement level in factors of 2 (up to a factor of 16) based on the local Courant condition \citep{Courant1928}. This increases the effective speed of the simulations several-fold while reducing the numerical noise \citep{Chatterjee2019}.

Since ideal GRMHD is unable to describe the physical processes responsible for injecting gas in the jet funnel  (such as pair creation) we use density floors in the drift frame \citep{Ressler2017} to ensure that ${\rho c^2}\ge{p_{mag}}/12.5$. However, due to the absence of any poloidal magnetic flux, no jets are launched in this work and thus the corresponding density floors are never activated. Instead, the ambient medium is floored at the density $\rho=10^{-7}\times r^{-2}$ and internal energy $u_{tot}=10^{-9} \times r^{-2.5}$, where $u_{tot}=u_{gas}+u_{rad}$  is the sum of the gas, $u_{gas}$, and radiation, $u_{rad}$, internal energies.

\begin{figure*}
    \centering
    \includegraphics[width=1.0\linewidth,trim=0mm 0mm 0mm 0,clip]{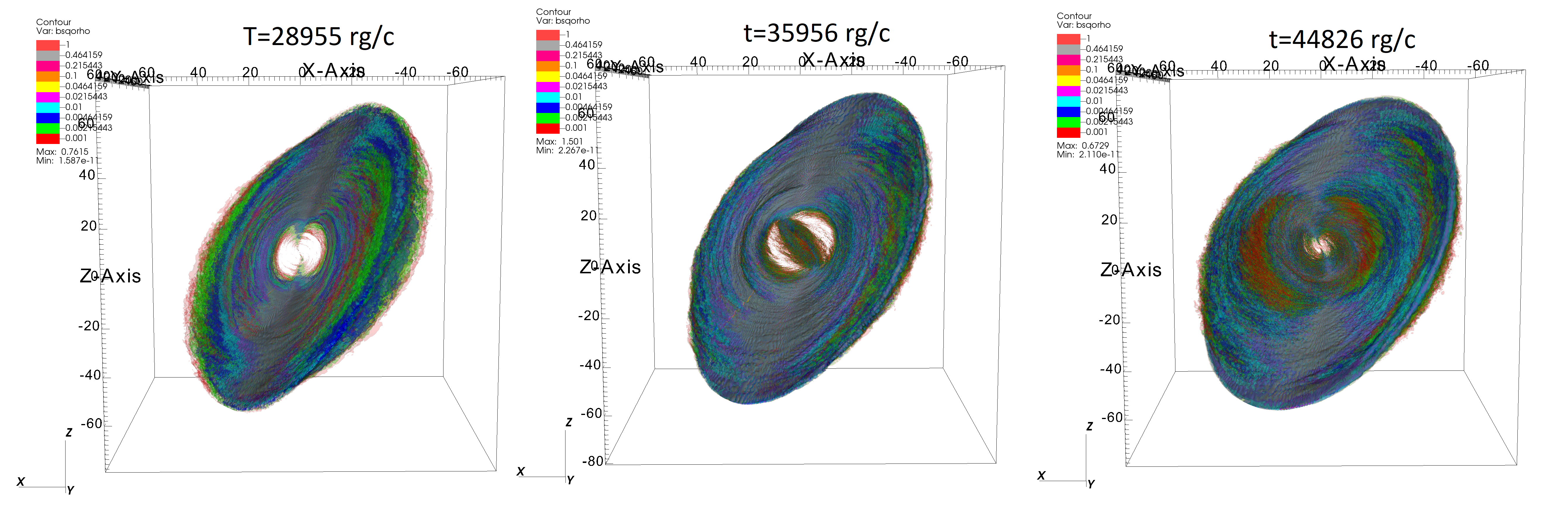}
    \vspace{-10pt}
    \caption{The first demonstration of disk tearing in a radiation pressure supported accretion disk, as seen through iso-countour 3D renderings of density at three different times. \textbf{(Left panel)} The disk at $t=28,955 r_g/c$ forms a rigid body that is warped, but shows no differential precession. \textbf{(Middle panel)} The disk tears off a precessing inner disk at $t=31,000 r_g$ that precesses for two periods between $t=35,000 r_g/c$ and $t=37,000 r_g/c$. While precessing, the inner disk aligns with the BH, shrinking in size until it fully disappears. \textbf{(Right panel)} The disk forms a new tear at $r \sim 25 r_g/c$ around $t \sim 45,000 r_g/c$, and the tearing process repeats. }
    \label{fig:contourplot_3d}
\end{figure*}

\section{Physical Setup}
\label{sec:Physical_Setup}
To study the effect of disk tearing in a highly warped accretion disk we evolve our model in two stages. In both stages we assume a rapidly spinning black hole of mass $M_{BH}=10M_{\odot}$ and dimensionless spin $a=0.9375$. The initial radial ($r$) and vertical ($z$) density profile of the disk is $\rho(r,z) \propto r^{-1}\exp{(-{z^2}/{2h^2})}$. The disk extends from an inner radius, $r_{in}=6 r_g$, to the outer radius, $r_{out}=76 r_g$ at a constant scale height $h/r = 0.02$. The covariant magnetic vector potential is $A_{\theta} \propto (\rho-0.0005)r^{2}$ and normalized to give an approximately uniform $\beta=p_{b}/p_{tot} \sim 7$ in the initial conditions. Here $p_b$ is the magnetic pressure and $p_{tot}$ is the sum of the electron/ion pressures, $p_{e,i}$, and the radiation pressure, $p_{rad}$. This rather low value of $\beta \sim 7$ was chosen such that the grid is able to resolve both the toroidal and poloidal MRI wavelengths for a physically meaningful runtime. A parameter exploration considering higher $\beta$ values will become possible in the coming $5-10$ years after the advent of (post-)exascale GPU clusters. The whole setup is subsequently rotated by 65 degrees with respect to the black hole spin axis, which itself is aligned with polar axis of the grid. Physical equivalence between tilting the disk (e.g. \citealt{Liska2018A, White2019A}) and tilting the black hole (e.g. \citealt{Fragile2005, Fragile2007}) is demonstrated in the Appendix of \citet{Liska2018A}.

In the first stage (model T65), which is described in \citet{Liska2020}, we evolve the disk in ideal GRMHD for $\Delta t = 150,000 r_g/c$ with a preset scale height of $h/r=0.02$ set by a cooling function \citep{Noble2009}. The disk undergoes a large tearing event around $t \sim 45,000 r_g/c$, which is analysed in \citet{Musoke2022}. During this tearing event a sub-disk of size $\Delta r \sim 10-15 r_g$ tears off and precesses for $\sim 6$ periods between $t  \sim 45,000r_g/c$ and $t \sim 80,000 r_g/c$. While precessing, the inner disk exhibits a very prominent radial epicyclic oscillation at $\nu = 56$~Hz for a 10 solar mass BH. Another tearing event between $t \sim 120,000 r_g/c$ and $t \sim 145,000 r_g/c$ produces a similar oscillation albeit at a slightly higher frequency of $\nu = 69$~Hz. These frequencies are consistent with observed high frequency QPOs (e.g. \citealt{IngramMotta2020}).

In the second stage (model RADT65), which is the focus of this work, we restart this simulation in full radiative GRMHD at $t=29,889 r_g/c$ as described above. The resolution is downscaled by a factor of $2$ to reduce the cost of the simulation and make it amenable to the pre-exascale GPU cluster OLCF Summit on which this simulation has been run. This is well before the occurrence of any large tearing event in stage 1. We assume thermodynamic equilibrium between gas and radiation by setting $T_r=T_e=T_i$ in the initial conditions and run this model (RADT65) for $\Delta t \sim 21,000 r_g/c$ until $t \sim 49,787 rg/c$. The density scaling is set such that the average accretion rate corresponds to approximately $\dot{M}/\dot{M}_{\rm Edd} \sim 0.35$ with $M_{\rm Edd}=\frac{1}{\eta_{NT}} L_{\rm Edd}/c^2$ the Eddington accretion rate, $L_{\rm Edd}=\frac{4 \pi G M_{BH}}{c k_{es}}$ the Eddington luminosity, and $\eta_{NT}=0.179$ the \citet{Novikov1973} efficiency. According to \citet{Piran2015} this choice of parameters is roughly consistent with a thin disk solution \citep{Sadowski2011} of scale height of $h/r=0.02-0.06$ between $r=r_{isco}$ and $r=100 r_g$. 

\begin{figure*}
    \centering
    \includegraphics[width=1.0\linewidth,trim=0mm 0mm 0mm 0,clip]{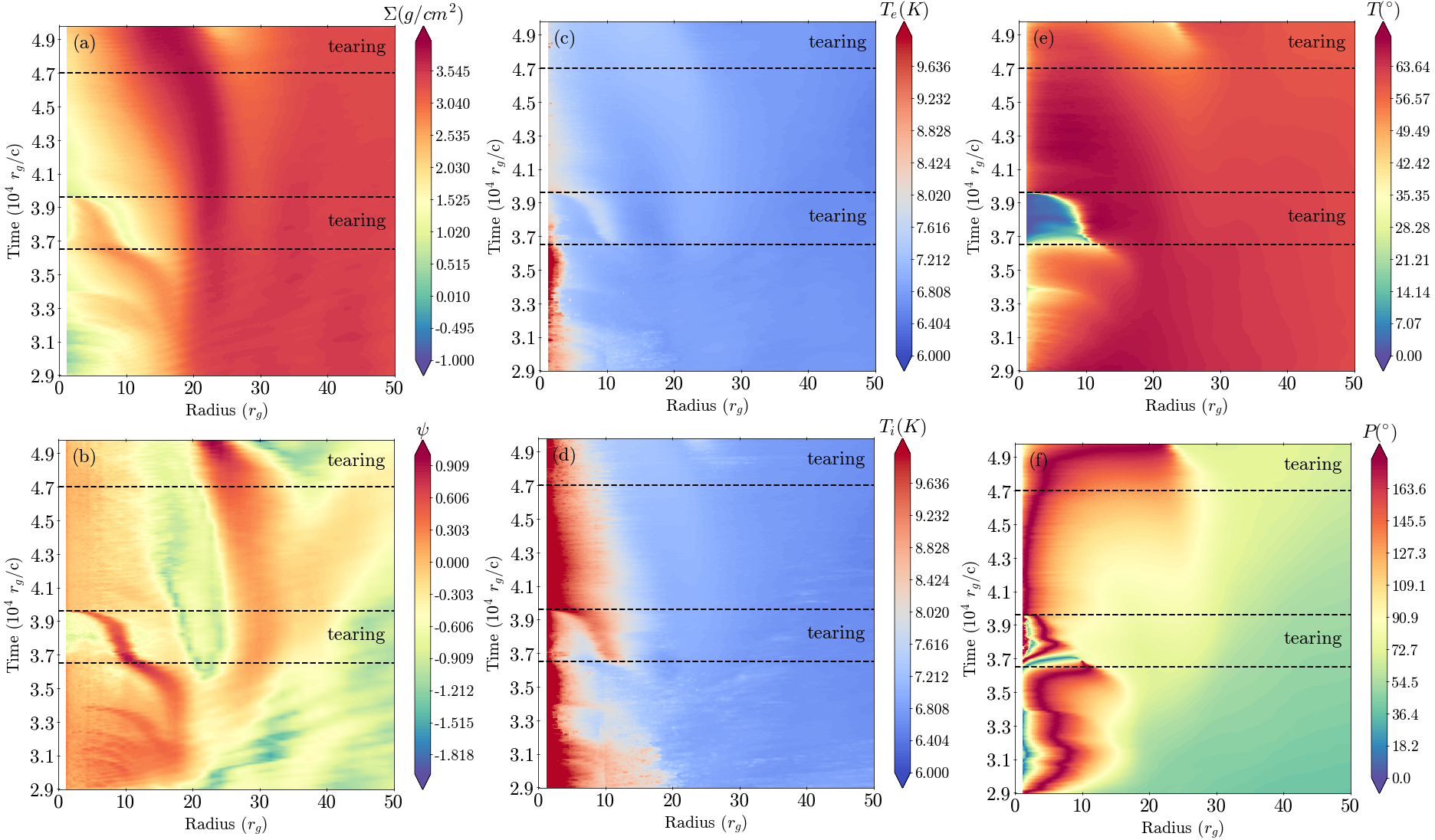}
    \caption{ \textbf{Panels (a, b)} Space-time diagrams of density, $\rho$, and warp amplitude, $\psi$. The warp amplitude peaks where the disk is about to tear or has already torn. Since dissipation is enhanced in warps, maxima in warp amplitude correspond to minima in density. The main tearing event is denoted by black dotted lines.  \textbf{Panels (c, d)} Space-time diagrams of the electron, $T_e$, and ion, $T_i$, temperatures. The ion and electron temperature correlate with the warp amplitude and anti-correlate with the density. Close to the black hole and in the tear the ions are significantly hotter than the electrons since Coulomb collisions are unable to equilibrate the ion and eletron temperatures. \textbf{Panels (e, f)} Space-time diagrams of the tilt, $\mathcal{T}$, and precession, $\mathcal{P}$, angle of the disk. Around $t \sim 36,500 r_g/c$ a disk tears off and precesses for $\lesssim 2$ periods, during which it gradually aligns with the black hole spin axis. Another tearing event occurs after $t \sim 47,000 r_g/c$. }
    \label{fig:ST}
\end{figure*}

\section{Results}
\label{sec:Results}
\subsection{Tearing Process}
\label{sec:Tearing}
The disk in RADT65 undergoes a tearing event between $t=31,000 r_g/c$ and $t=40,000 r_g/c$, as visualised using a 3D density isocontour rendering at three different times in Fig.~\ref{fig:contourplot_3d}). Figure~\ref{fig:ST} shows the space-time diagram, in which we calculate the average density of the disk $\bar{\rho}=\frac{\int^{2 \pi}_{0} \int^{\pi}_{0} \rho^{2} \sqrt{-g} d\theta d\phi}{\int^{2 \pi}_{0} \int^{\pi}_{0} \rho \sqrt{-g} d\theta d\phi}$ (Fig.~\ref{fig:ST}a), the warp amplitude, $\Psi=r \frac{\partial \vec{l}}{\partial r}$ with $\vec{l}$ being the angular momentum (Fig.~\ref{fig:ST}b), the average electron and ion temperatures, $T_{e,i}=\frac{\int^{2 \pi}_{0} \int_{0}^{\pi} p_{e,i} \sqrt{-g} d\theta d\phi}{\int^{2 \pi}_{0} \int^{\pi}_{0} \rho \sqrt{-g} d\theta d\phi}$ (Fig.~\ref{fig:ST}c,d), the tilt and precession angles, $T_{disk}$ and $P_{disk}$ (Fig.~\ref{fig:ST}e,f as calculated in \citealt{Fragile2005}). 

During this tearing event the disk precesses for $\lesssim 2$ periods between $36,000<t<38,000 r_g/c$ (Fig.~\ref{fig:ST}e) before it aligns with the spin axis of the black hole (Fig.~\ref{fig:ST}f). Alignment happens through warp driven dissipation of misaligned angular momentum on the accretion time of the inner disk, which differs from the Bardeen-Petterson alignment \citep{bp75} mechanism that manifests itself on much shorter timescales during which the structure and density of the disk remains in a steady state (see discussion in \citealt{Liska2018C}). In addition to the first tearing event, which will be the focus of this article, a much larger tearing event starts after $t \gtrsim 47,000 r_g/c$ that does not finish before the end of this simulation. Due to the large computational expense associated with extending the duration of RADT65, we leave the full analysis of this longer tearing event to future work.

Whether a disk is able to tear depends on how the internal torques in the disk react to differential Lense-Thirring precession \citep[e.g.][]{Nixon2012}. The Lense-Thirring precession rate of a point particle around a rapidly spinning black hole follows a steep radial dependence of $\nu_{LT} \propto a/r^3$. However, an accretion disk will precess with a single frequency set by the integrated LT torque in the disk if the viscous torques are able to counteract the differential precession rate (e.g. \citealt{Fragile2005, Fragile2007}). Naively, one expects that the warp amplitude decreases monotonically with the increasing distance from the black hole. This is (roughly) the case in the inner and outer disk of RADT65 for $t \lesssim 32,000 r_g/c$. 

As Fig.~\ref{fig:ST}(a),(b) shows, the warp amplitude and surface density in RADT65 are anti-correlated for the entire runtime. At $r \lesssim 20 r_g$ the warp amplitude is rather high and a cavity of low density gas forms at both early and late times. In this cavity the plasma thermally decouples into hot electrons with $T_e \sim 10^8-10^9K$ and extremely hot ions with $T_i \sim 10^9-10^{10} K$. The strong anti-correlation between warp amplitude and density suggests that accretion is driven by dissipation in warps (e.g. \citealt{Papaloizou1983, Ogilvie1999, Nelson2000}). In a companion paper \citep{Kaaz2022} we demonstrate that warp-driven dissipation is responsible for the bulk of the dissipation and that magneto-rotational instability driven turbulence (e.g. \citealt{Balbus1991, Balbus1998}) plays only a very subdominant role. 

However, after $t \gtrsim 32,000 r_g/c$ the warp amplitude is not a monotonically decreasing function of radius. Instead, the warp amplitude forms local minima and maxima, and concentric rings of higher density gas centered at the (local) minima of the warp amplitude form. It is known that when the warp amplitude exceeds a critical value, it can become unstable \citep{Dogan2018, Dogan2020}. When this happens, the viscous torques in the warp drop to zero and the disk tears. In RADT65 the tearing process starts at $r \sim 19 r_g$ around $33,000 r_g/c$ and the disk fully detaches at $r \sim 10 r_g$ around $t \sim 36,500 r_g/c$. We see later that during this process warp-driven dissipation leads to a temporary burst in the BH accretion rate and luminosity around $t \sim 36,500 r_g/c$ (Sec.~\ref{sec:variability}). The disk even (briefly) enters the slim disk regime where $\dot{M}\sim \dot{M}_{edd}$ and $L \sim 0.5 L_{edd}$, suggesting the luminosity is slightly suppressed due to photon trapping (e.g. \citealt{Abramowicz1978}). Similarly, we observe for the second tearing event after $t\sim 47,000 r_g/c$ that $\dot{M}\sim 3\dot{M}_{edd}$ and $L \sim 1.0 L_{edd}$.

\begin{figure}
    \centering
    \includegraphics[width=\columnwidth,trim=0mm 0mm 0mm 0,clip]{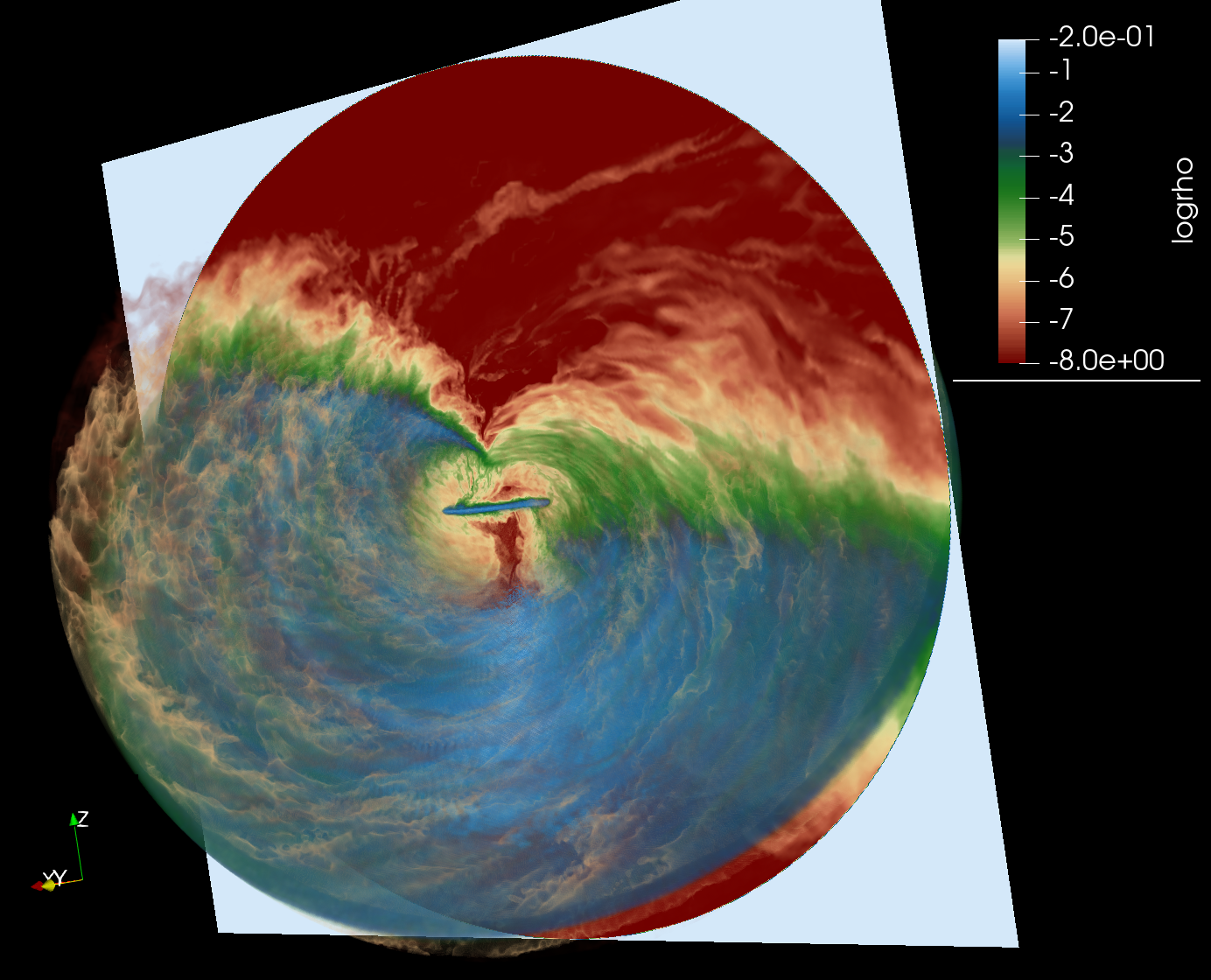}
    \vspace{-10pt}
    \caption{ A volume rendering of the density during the main tearing event at $t=37,026 r_g/c$. The red plane cuts the disk at the $\phi=0$ surface. Streamers from the outer disk (green) deposit low angular momentum gas onto the inner disk (blue). This causes the inner accretion disk to shrink in size \citep{Kaaz2022}.}
    \label{fig:streamers}
\end{figure}

As the gas from the inner disk falls into the black hole it transports angular momentum outwards. Angular momentum conservation dictates that the radius of the inner disk should increase. This process, called viscous spreading, affects all finite size accretion disks that are not resupplied externally with gas (e.g. \citealt{Liska2018A, Porth2019}). However, the inner disk in RADT65 does not spread viscously, but, in fact, decreases in size. This discrepancy could potentially be explained by cancellation of misaligned angular momentum (e.g. \citealt{Nixon2012, Hawley2019}). Namely, when gas from the outer disk falls onto the inner disk the misaligned components of angular momentum cancel out leading to a net decrease in the inner disk radius. In addition, it is possible that the inner disk transfers some of its excess angular momentum to the outer disk. In a companion paper \citep{Kaaz2022} we quantify the relative contributions of the angular momentum cancellation and outward angular momentum transport to the evolution of the tearing radius.

\begin{figure*}
    \centering
    \includegraphics[width=\linewidth,trim=0mm 0mm 0mm 0,clip]{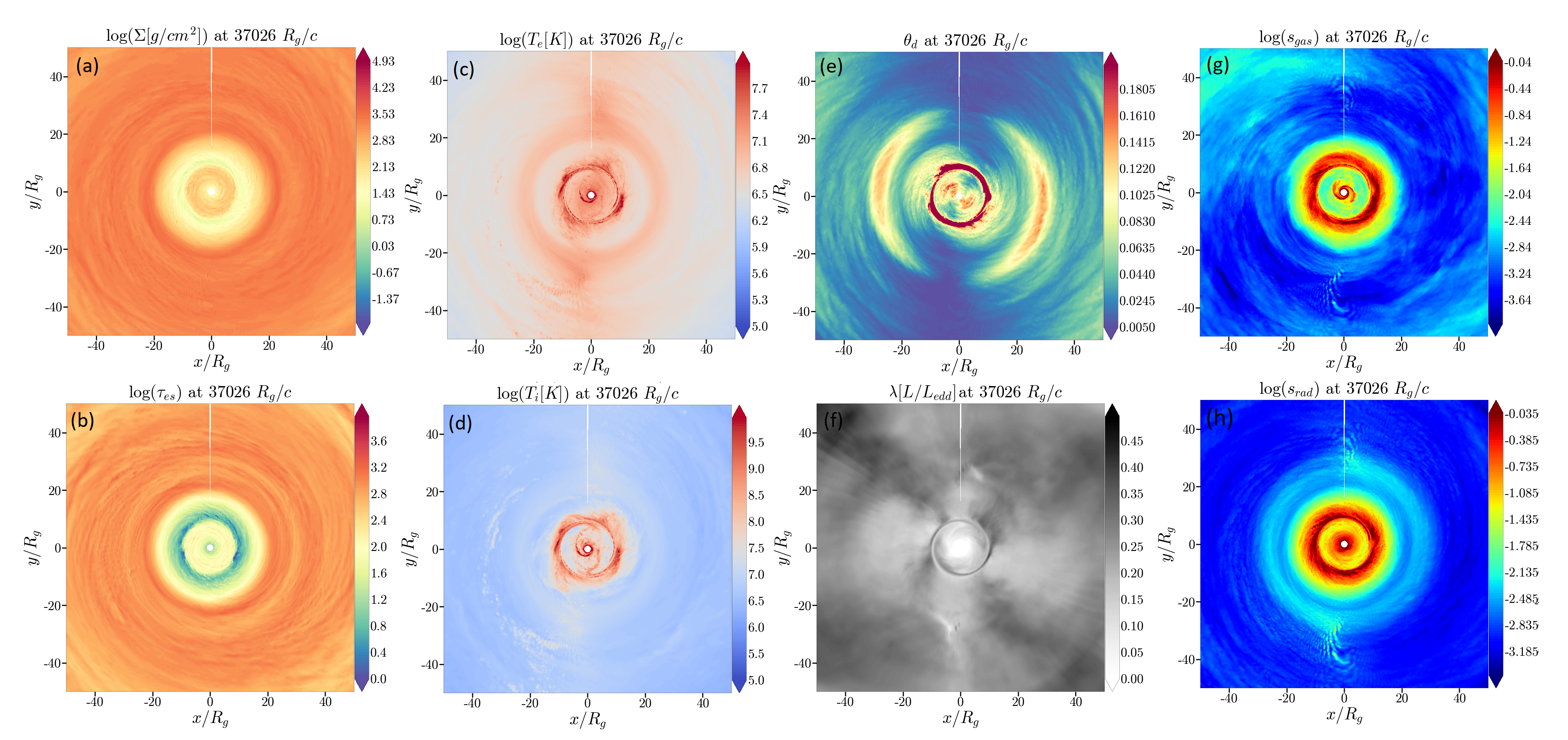}
    \vspace{-10pt}
    \caption{A top down view of the accretion disk during the main tearing event at $t=37,026 r_g/c$ corresponding to the volume rendering in Figure \ref{fig:streamers}. \textbf{(Panel a)} The surface density, $\Sigma$, drops sharply in the tear due to additional dissipation as the gas undergoes a radical plane change. \textbf{(Panel b)} The vertically integrated scattering optical depth ($\tau_{es}$) drops below unity in the tear suggesting optically thin emission. \textbf{(Panels c, d)} The electron, $T_e$, and ion, $T_i$, temperatures peak perpendicular to the line of nodes (due to a `nozzle' shock, see Sec.~\ref{sec:rad_sign}) and in the tear at $r \sim 10 r_g$. \textbf{(Panel e)} The density scale height, $\theta_d$, of the accretion disk oscillates in azimuth. It peaks along the x-axis and reaches a minimum along the y-axis. \textbf{(Panel f)} The bolometric luminosity, $\lambda$, forms a non-axisymmetric pattern ranging from $\lambda \sim 0.15 L_{\rm Edd}$ to $\lambda \sim 0.5 L_{\rm Edd}$. \textbf{(Panels g, h)} The gas entropy entropy, $s_{gas}$, does not increase substantially when gas crosses the scaleheight compression along the vertical y-axis, but the radiation entropy, $s_{rad}$, does. This suggests the shock heating is radiatively efficient.}
    \label{fig:vertical_plot}
\end{figure*}

\subsection{Dissipation and Radiative Signatures}
\label{sec:rad_sign}
To better understand the structure of the disk during the tearing event at $t=37,026 r_g/c$ we show a volume rendering of the density in Figure~\ref{fig:streamers} and a vertical projection of several quantities in Figure \ref{fig:vertical_plot}. In this projection the line of nodes, where the disk crosses the equatorial plane, is aligned with the horizontal x-axis. The projected quantities include the surface density $\Sigma=\int_{0}^{\pi} \rho \sqrt{g_{\theta \theta}} d\theta$ (panel a),  the scattering optical depth $\tau_{es} \sim 0.34 \Sigma$ (panel b), electron and ion temperature $T_{e,i}=\frac{\int_{0}^{\pi}p_{e,i}\sqrt{g_{\theta \theta}}d\theta}{\Sigma}$ (panels c and d), density scale height $\theta_{d}=\frac{\int_{0}^{\pi} \rho(\theta - \pi/2) \sqrt{g_{\theta \theta}}d\theta}{\Sigma}$ (panel e), effective bolometric luminosity $\lambda=\frac{\int_{0}^{\pi} R^{r}_{t} \sqrt{g_{\theta \theta}} 2 \pi d\theta}{L_{edd}}$ (panel f), gas entropy per unit mass $s_{gas}=\frac{\int_{0}^{\pi}p_{e,i} /\rho^{\gamma} \rho u^{t} \sqrt{g_{\theta \theta}}d\theta}{\Sigma}$ (panel e), and radiation entropy per unit mass  $s_{rad}=\frac{\int_{0}^{\pi} E_{rad}^{3/4} \sqrt{g_{\theta \theta}}d\theta}{\Sigma}$ (panel h). Integration is performed in a spherical coordinate system aligned with the local angular momentum vector of the disk. An animation of Figure \ref{fig:streamers} and panels a, c and f of Figure \ref{fig:vertical_plot} can be found on our YouTube channel (\href{https://www.youtube.com/watch?v=JtM3XDIUsro}{movie}).

Interestingly, the electron and ion temperatures do not decline smoothly with distance from the black hole (Fig.~\ref{fig:vertical_plot}b,c) as predicted by idealized models of thin accretion disks \citet{Novikov1973}. At the tearing radius of $10 r_g$, where the warp amplitude reaches a maximum, the electron temperature reaches $T_e \gtrsim 10^8K$ and the plasma becomes optically thin. Here streamers of gas get thrown onto highly eccentric orbits and subsequently rain down on the inner disk (Fig.~\ref{fig:streamers}). The rapid rise in temperature in the tear suggests that gas crossing the tear is subject to additional dissipation as it undergoes a rapid orbital plane change (see also \citealt{Nixon2014, Raj2021_AGN}). If we assume the gas has a temperature of $T_i=0K$ before crossing the tear at $r \sim 7 r_g$, and the gas dissipates $\epsilon=10^{-2}$ of its orbital kinetic energy $U_{kin}\approx\frac{1}{2r}$, we find that dissipation can easily heat the gas to $T_i \sim 10^{10} K$. In reality, $\epsilon$ can be much higher if radiative cooling is efficient.

\begin{figure}
    \centering
    \includegraphics[clip,trim=0.0cm 0.0cm 0.0cm 0.0cm,width=\columnwidth]{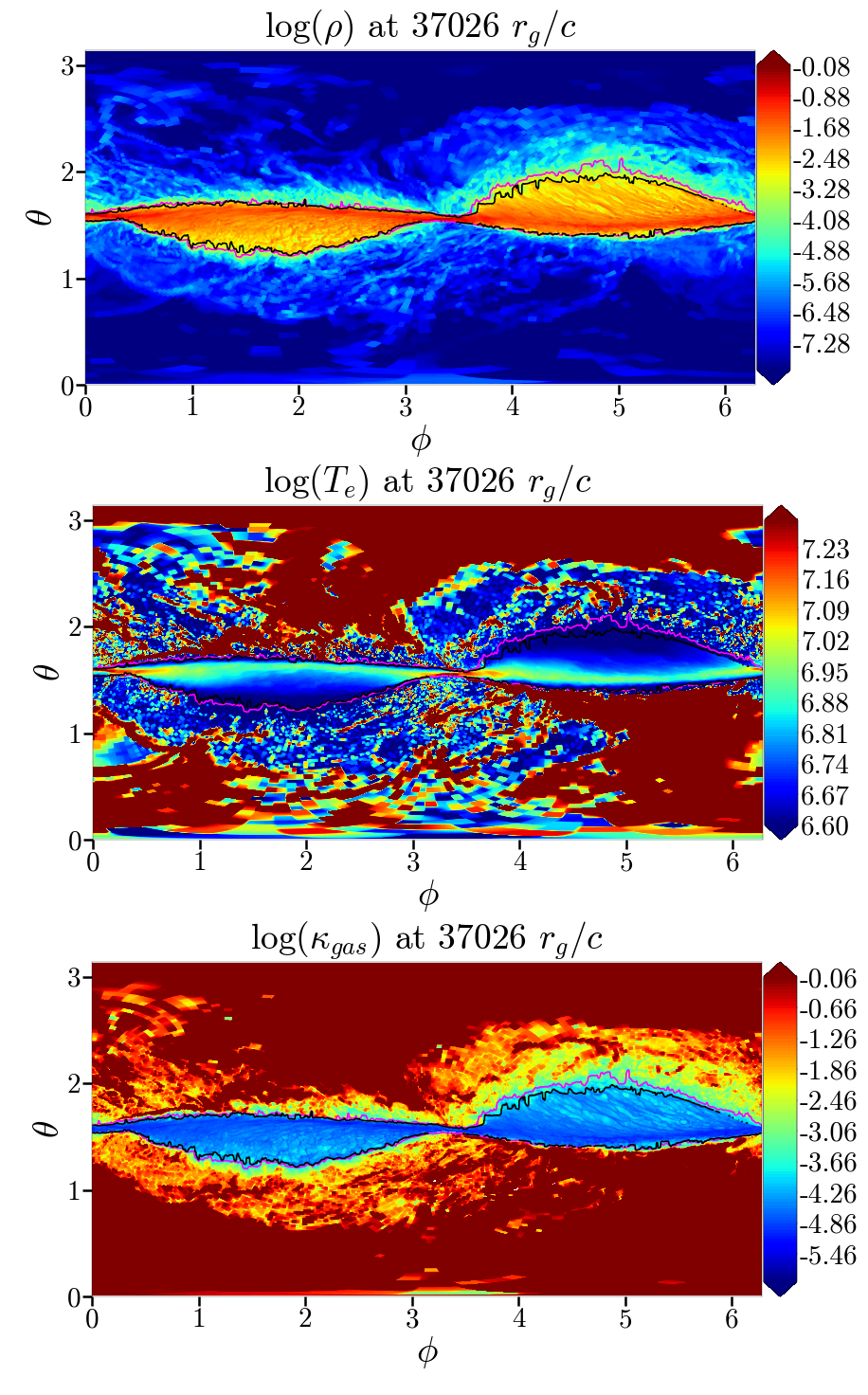}
    \vspace{-10pt}
    \caption{A $\theta-\phi$ slice through the disk at $r=15 r_g$ in tilted coordinates with the ascending node at $\phi=1/2\pi$ and the descending node at $\phi=3/2\pi$. The pink and black lines correspond to the $\tau_{es}=1$ and $\tau_{bf}=1$ surfaces respectively. \textbf{(Upper panel)} The gas density, $\rho$, peaks in the `nozzle', which is located perpendicular to the local line of nodes at $\phi=0$ and $\phi=\pi$. \textbf{(Middle panel)} The electron temperature, $T_e$, increases by a factor $\sim 5$ in the nozzle, suggesting the radiative emission will be hardened. \textbf{(Lower panel)} The gas entropy per unit mass, $\kappa_{gas}$, does not increase in the nozzle, suggesting the rise in gas temperature is adiabatic (Sec. \ref{sec:rad_sign})}
    \label{fig:phiplot}
\end{figure}

The scale height of the disk exhibits a prominent $m=2$ azimuthal oscillation  (Fig.~\ref{fig:vertical_plot}e). This can also be observed in Figure \ref{fig:phiplot} where we show a $\theta-\phi$ slice at $r=25 r_g$ of the density, electron temperature, and gas entropy. At the line of nodes the scale height reaches a maximum of $h/r \sim 0.1$ while perpendicular to the line of nodes the scale height drops to $h/r \sim 0.005$. Around this nozzle-like `compression', the gas reaches a temperature ranging from $T_e \sim 2 \times 10^7 K$ at $r \sim 20 r_g$ to $T_e \gtrsim 10^8K$ within $r \lesssim 5 r_g$. At select times $T_e$ can even reach $T_e \sim 10^9K$ closer to the black hole (see YouTube \href{https://www.youtube.com/watch?v=JtM3XDIUsro}{movie}). The radial and azimuthal fluctuations in the temperature lead to a highly non-axisymmetric emission pattern (Fig.~\ref{fig:vertical_plot}f). At this specific snapshot ($t=37,026 r_g/c$) the effective bolometric luminosity varies azimuthally between $\lambda \sim 0.15 L_{edd}$ and $\lambda \sim 0.5 L_{edd}$ at $r=200 r_g$. 

The specific gas entropy does not increase in the nozzle (Fig.~\ref{fig:vertical_plot}g), suggesting heating of gas in the nozzle is adiabatic. This might seem surprising since nozzles in tidal disruption events (TDEs) are typically associated with shock heating and steep gradients in the gas entropy (e.g. \citealt{Rees1988, Kochanek1994, Andalman2020}). However, if the radiative cooling timescale is very short, shocks do not automatically lead to an increase of the gas entropy. Instead, they will lead to an increase in the radiation entropy. This increase in radiation entropy is visible along the vertical y-axis in Fig.~\ref{fig:vertical_plot}(g). In a companion paper \citep{Kaaz2022}, we demonstrate that a shock forms in the nozzle that dissipates roughly $\sim 1.8\%$ of the orbital energy each time the gas passes through it.

\subsection{Outflows and Variability}
\label{sec:variability}
\begin{figure}
    \centering
    \includegraphics[clip,trim=0.0cm 0.0cm 0.0cm 0.0cm,width=\columnwidth]{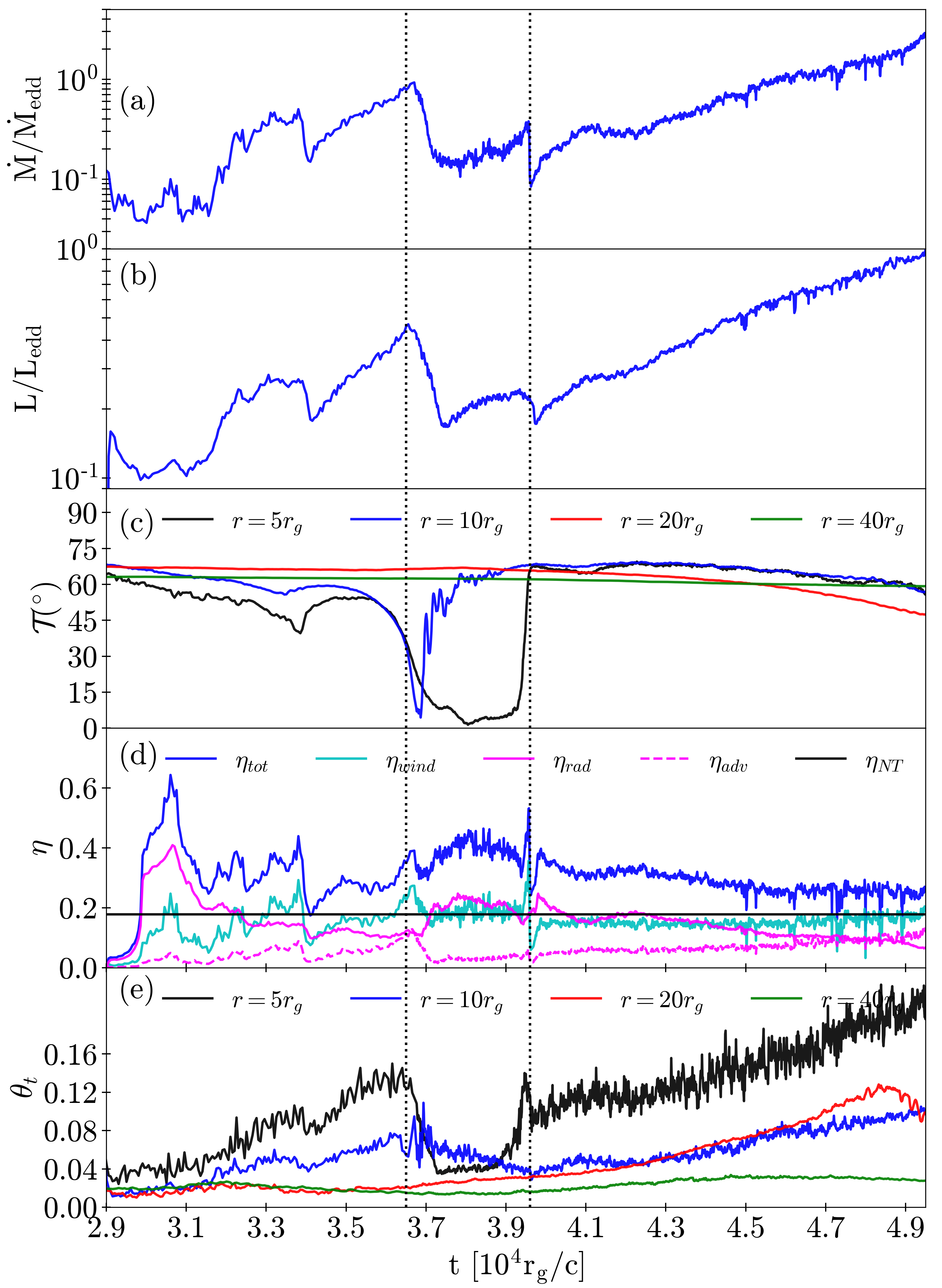}
    \vspace{-10pt}
    \caption{Time evolution of RADT65 with the main tearing event denoted by black dotted lines. \textbf{(Panel a)} The mass accretion rate increases from $\dot{M} \sim 10^{-2} \dot{M}_{\rm Edd}$ to  $\dot{M} \sim 3 \dot{M}_{\rm Edd}$. \textbf{(Panel b)} The luminosity increases from $L \sim 0.1 L_{\rm Edd}$ to $L \sim 1.0 L_{\rm Edd}$. \textbf{(Panel c)} The tilt $\mathcal{T}$ of the accretion disk. The inner accretion disk ($r \lesssim 10 r_g$) aligns with the black hole just after the main tearing event while the outer accretion disk remains misaligned. \textbf{(Panel d)} The wind efficiency, $\eta_{wind}$, radiative efficiency, $\eta_{rad}$, advective radiative efficiency, $\eta_{adv}$, and \citet{Novikov1973} efficiency, $\eta_{NT}$, and total efficiency, $\eta_{tot}$, as defined in Sec.~\ref{sec:variability}. When the accretion rate peaks, advection of radiation becomes important and suppresses the total luminosity of the system. \textbf{(Panel e)} The thermal scale height of the disk, $\theta_{t}$, stays stable or increases throughout the disk, which suggests the disk is thermally stable (Sec.~\ref{sec:stability}).}
    \label{fig:timeplot}
\end{figure}

To understand the variability associated with disk tearing and the energetics of the outflows we plot in Figure~\ref{fig:timeplot} the mass accretion rate (panel a), radiative emission rate (panel b) and (radiative) outflow efficiencies (panel c).  The mass accretion rate measured at the event horizon ($\dot{M}=\int_{0}^{2\pi} \int_{0}^{\pi} \rho u^r \sqrt{-g} d\theta d\phi$ with $g$ the metric determinant and $u^{\mu}$ the fluid 4-velocity) exhibits a factor $\sim 40$ variation in time, ranging from $\dot{M}=2 \times 10^{-2} \dot{M}_{\rm Edd}$ to $\dot{M}=\dot{M}_{\rm Edd}$ (defined in Sec.~\ref{sec:Physical_Setup}). However, the luminosity ($L=\int_{0}^{2\pi} \int_{0}^{\pi} R^{r}_{t} \sqrt{-g} d\theta d\phi$ with $R^{r}_{t}$ the $r$-$t$ component of the radiation stress energy tensor) only exhibits a factor $\sim 5$ variation and peaks at $L \sim 0.5 L_{edd}$. 

The outflow efficiencies with respect to the BH mass accretion rate are defined at the event horizon as $\eta_{wind}=\frac{\dot{M}-\dot{E}}{|\dot{M}|_{t}}$, $\eta_{rad}=\frac{L_{r=r_{200 r_g}}}{|\dot{M}|_t}$, and $\eta_{adv}=\frac{-L_{r=r_{BH}}}{|\dot{M}|_t}$. Here $\dot{E}=\int_{0}^{\pi} \int_{0}^{2\pi} T^{r}_{t} \sqrt{-g} d\theta d\phi$ is the energy accretion rate with $T^{r}_{t}$ the stress energy tensor, and, $|\dot{M}|_t$ is a running average of $\dot{M}$ over a time interval of $500 r_g/c$. Unless stated otherwise, we measure $\dot{M}$ and $\dot{E}$ at the event horizon, $L$ at the event horizon for $\eta_{adv}$ and at $r=200 r_g$ for $\eta_{rad}$. As Fig.~\ref{fig:timeplot}(c) shows, when the accretion rate reaches the Eddington limit the advective efficiency, $\eta_{adv}$, is of comparable magnitude to the radiative efficiency,  $\eta_{rad}$ (Fig.~\ref{fig:timeplot}d). This implies that a significant fraction of the emitted radiation will fall into the black hole before it reaches the observer suggesting the disk might have entered the slim disk regime at the peak of luminosity. Nevertheless, the time-averaged radiative efficiency reaches $|\eta_{rad}|=14.7 \%$, which is only slightly below the canonical \citet{Novikov1973} efficiency of $\eta_{NT}=17.8\%$.

We do not observe any evidence for a radiation or magnetic pressure driven wind. While the outflow efficiency measured at the event horizon exceeds $\eta_{wind}\gtrsim 15 \%$ this drops to $\eta_{wind} \lesssim 0.1 \%$ when measured at $r=200 r_g$, which demonstrates that both the magnetic pressure (e.g. \citealt{Liska2018C}) and radiation pressure (e.g. \citealt{Kitaki2021}) are insufficient to accelerate the wind to escape velocities. This suggests that poloidal magnetic fields are a key ingredient to accelerate sub-relativistic outflows, even when the accretion rate approaches the Eddington limit.

\subsection{Radial Structure}
\label{sec:rad_structure}
To better understand the structure of the accretion disk we show in Figure~\ref{fig:radplot} time-averaged (between $t=[43,000, 44,000] r_g/c$) radial profiles of the density, $\bar{\rho}$ (panel a), luminosity, $L$ (panel b), and radiation, ion, electron, and magnetic pressure, $p_{rad}$, $p_i$, $p_e$, and $p_{b}$, respectively (panel c). During this time interval a new tear starts to form around $r \sim 30 r_g$, which manifests itself as a drop in density and increase in the radiative flux. While the magnetic and radiation pressure are in approximate equipartition within $r \lesssim 20 r_g$, the accretion disk becomes fully radiation pressure dominated between $r=[20,40] r_g$. The sum of the ion and electron pressure is a factor $\sim 10-100$ lower than the radiation pressure, suggesting that gas pressure does not play a dynamically important role.

We have already mentioned in Section \ref{sec:Tearing} that warp driven dissipation plays an important role in RADT65. To illustrate how much warp-driven dissipation accelerates the inflow of gas in RADT65, we translate the radial infall speed of the gas into an effective viscosity, $\alpha_{\rm eff}=\frac{\langle v^{r}v^{\phi}\rangle_{\rho}}{\langle c_{s}^{2}\rangle_{\rho}}$ with $v^{r}=u^{r}/u^{t}$ and $v^{\phi}=u^{\phi}/u^{t}$. Here, $\langle x\rangle_y=\frac{\int_{0}^{2\pi}\int_{0}^{\pi} x y \sqrt{-g} d\theta d\phi d\phi}{\int_{0}^{2\pi}\int_{0}^{\pi} y \sqrt{-g} d\theta d\phi}$ gives the $y-$weighted average of quantity $x$. In Fig.~\ref{fig:radplot}(c), we find $\alpha_{\rm eff} \sim 1-5 \times 10^1$ throughout the disk. The Maxwell stresses $\alpha_{\rm m}=\frac{\langle b^{r}b^{\phi}\rangle_{\rho}}{\langle p_{tot}\rangle_{\rho}} \sim 10^{-2}-10^{-1}$ induced by magnetized turbulence cannot account for this effective viscosity, which suggests that other physical processes such as shocks (e.g. \citealt{Fragile2008}) drive accretion in warped accretion disks.

\subsection{Thermal Stability}
\label{sec:stability}
Interestingly, \citet{Liska2022} found that an aligned version of the accretion disk in RADT65 collapsed within $t \lesssim 5,000 r_g/c$ to an exceedingly thin slab that could not be resolved numerically. This runaway cooling (see also \citealt{Sadowski2016_mag,Fragile2018, Jiang2019, Mishra2022}) is a manifestation of the thermal instability \citep{Shakura1976}. However, we do not observe any evidence for the thermal instability in RADT65. In fact, as Fig.~\ref{fig:timeplot}(e) demonstrates, the thermal scale height, $\theta_{t}=\frac{\langle c_s\rangle_{\rho}}{\langle v^{\phi}\rangle_{\rho}}$, remains stable or even increases depending on the radius. 

This unexpected result could potentially be explained within $r\lesssim 20 r_g$ by magnetic fields, which can stabilize a thermally unstable accretion disk if they reach equipartition with $p_b \sim p_{rad}$  (e.g.  \citealt{Begelman2007, Sadowski2016_mag, Jiang2019, Liska2022, Mishra2022}). However, magnetic fields will not be able to stabilize the disk between $r=[20,40] r_g$ where the disk is strongly radiation pressure dominated with $p_{rad}/p_{b} \sim 10^1-10^2$. 

\begin{figure}
    \centering
    \includegraphics[clip,trim=0.0cm 0.0cm 0.0cm 0.0cm,width=\columnwidth]{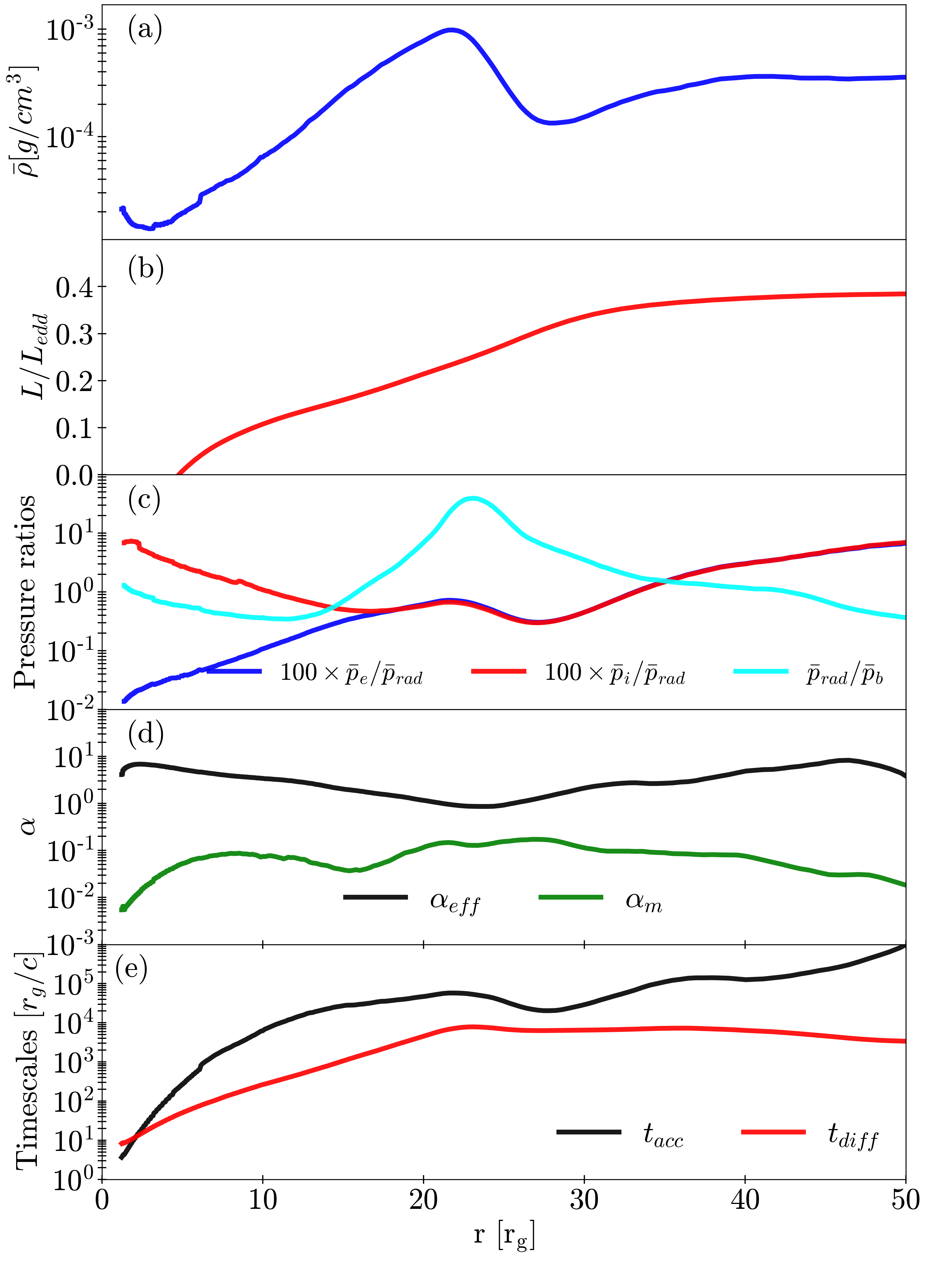}
    \vspace{-10pt}
    \caption{Radial structure of the disk time-averaged between $t=[43,000; 44,000] r_g/c$. This is the end state of our simulation at which point a new tear starts to form. \textbf{(Panel a)} The gas density, $\bar{\rho}$, peaks around $r \sim 25 r_g$. We expect the new tear to form just behind the peak in density. \textbf{(Panel b)} The cumulative luminosity of the disk increases until $35 r_g$, partially driven by enhanced dissipation in the forming tear between $r=20 r_g$ and $r=30 r_g$. \textbf{(Panel c)} The disk is radiation pressure, $p_{rad}$, dominated between $r=[15, 40] r_g$ and magnetic pressure, $p_b$, dominated everywhere else. The ion, $p_i$), and electron, $p_e$, pressures are very weak compared to either the radiation or magnetic pressure and thus do not play an dynamically important role. \textbf{(Panel d)} The effective viscosity, $\alpha_{\rm eff}$, is much larger than the Maxwell, $\alpha_{\rm m}$, stress. This suggests that dissipation is not driven by magnetized turbulence induced by magnetic stresses. Instead, we demonstrate in a companion paper that shocks are driving accretion \citep{Kaaz2022}. \textbf{(Panel e)} The photon diffusion timescale, $t_{\rm diff}$, is shorter than the accretion timescale, $t_{\rm acc}$, except very close to the black hole. This suggest advection of radiation internal energy is only dynamically important very close to the black hole. }
    \label{fig:radplot}
\end{figure}

We found in Sec.~\ref{sec:variability} that advection of radiation becomes significant at the event horizon. If the radiation diffusion timescale becomes longer than the accretion timescale, photon trapping might stabilize the disk against runaway cooling (e.g. \citealt{Abramowicz1978}). To test if photon trapping can explain the apparent thermal stability of the disk we plot the radiation diffusion time, $t_{\rm diff}\sim \frac{h}{v_{rad}} \sim 3 h \tau_{es}$ and the accretion time, $t_{\rm acc}=\frac{r}{\langle v_r\rangle_{\rho}}$, in Fig.~\ref{fig:radplot}(e). Here we use for $v_{rad}$ the optically thick radiation diffusion speed $v_{\rm rad} \sim \frac{1}{3 \tau_{es}} c$ (e.g. \citealt{Mckinney2013}). As is evident from Fig.~\ref{fig:radplot}(e), while the radiation diffusion and accretion timescales are comparable in the inner disk, the radiation diffusion timescale becomes much shorter than the accretion timescale in the outer disk. This suggests that photon trapping will not be able to stabilize the outer disk against runaway cooling or heating. 

Instead, we propose in Section \ref{sec:discussion} that nozzle shock driven accretion disks are inherently thermally stable.

\section{Discussion}
\label{sec:discussion}
\subsection{Applicability to Spectral States}
The accretion disk in RADT65 does not launch any relativistic jet. This suggests RADT65 is applicable to the high-soft and potentially soft-intermediate spectral states. In these spectral states the spectrum is dominated by thermal blackbody radiation and no radio jets are detected (though see \citealt{Russel2020} for detection of compact jets in the infrared).

Nevertheless, the emission profile of RADT65 presents a radical departure from the \citet{Novikov1973} model of a thin accretion disk. Instead of an axisymmetric power-law emission profile, emission in warped accretion disks can become hardened around the nozzle shock or at the tearing radius (see also \citealt{Nixon2014, Raj2021_AGN}). Here the temperature of plasma can reach $T_e \sim 10^8-10^9 K$, which might lead to Comptonization of cold accretion disk photons. This could potentially contribute to the high-energy power law emissions tail observed in the spectra of soft state BHXRBS (e.g. \citealt{Remillard2006}) in addition to the soft X-Ray excess in the spectra of high-luminosity AGN (e.g. \citealt{Gierlinski2004, Gierlinski2006}). Note that part of this high-energy emission can come from within the ISCO as suggested by recent GRMHD simulations \citep{Zhu2012} and semi-analytical models \citep{Hankla2022}.  Future general relativistic ray-tracing (GRRT) calculations will need to quantify how much the spectrum deviates from a blackbody spectrum and in which wavelength bands this optically thin shock-heated gas can be detected. The energy spectrum  of the accelerated electrons will depend on the microphysics of the shocks, which will need to be addressed by particle-in-cell simulations.

In the hard-intermediate state (HIMS) jetted ejections imply that the accretion disks might be saturated by vertical magnetic flux (e.g. \citealt{Mckinney2012, Tchekhovskoy2011, Chatterjee2020}). When an accretion disk is saturated by magnetic flux it enters the magnetically arrested disk (MAD) regime \citep{Narayan2003} in which magnetic pressure overcomes gravity and accretion proceeds through non-axisymmetric instabilities. Recent two-temperature radiative GRMHD simulations \citep{Liska2022} have demonstrated that magnetic flux saturation can lead to `magnetic' truncation during which the inner disk decouples into a magnetic pressure supported `corona' of hot electrons ($T_e \gtrsim 5 \times 10^8 K$) and extremely hot ions ($T_i \gtrsim 10^{10}K$). This happens within the magnetic truncation radius, which is determined by the amount and location of the excess poloidal magnetic flux in the system. It is conceivable that the combined effects of magnetic reconnection in current sheets and shocks in warps can heat the plasma to temperatures far above what was observed in this work or in \citet{Liska2022}. Future work will need to address if, where and how such misaligned truncated disks tear, and, if their spectral and timing signatures are consistent with sources in the HIMS.

\subsection{Quasi-periodic Variability}
\begin{figure}
    \centering
    \includegraphics[clip,trim=0.0cm 0.0cm 0.0cm 0.0cm,width=\columnwidth]{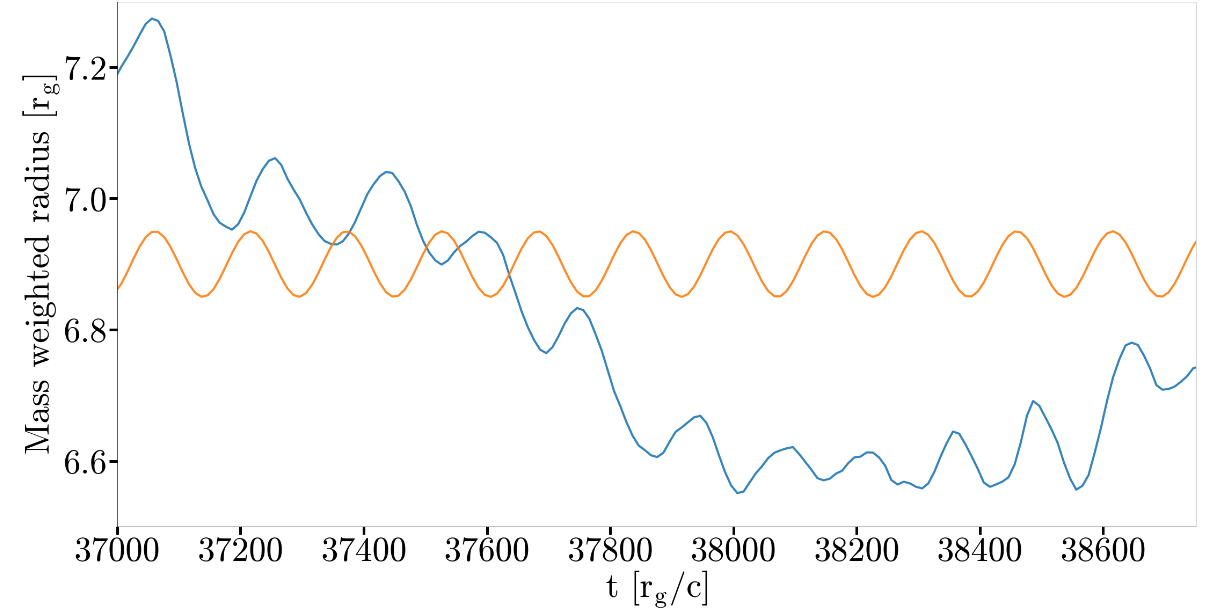}
    \vspace{-10pt}
    \caption{The mass weighted radius, $r_{M}=\frac{\int_{0}^{2 \pi} \int_{0}^{\pi} \Sigma r^3 \sqrt{-g} d\theta d\phi}{\int_{0}^{2 \pi} \int_{0}^{\pi} \Sigma r^2 \sqrt{-g} d\theta d\phi}$, of the inner disk (after tearing) in RADT65 (blue) features a radial epicyclic oscillation consistent with the radial epicyclic frequency of an oscillating ring of gas at $r=6.9 r_g$ (orange). Such oscillations may be associated with HFQPOs detected in BHXRB lightcurves.}
    \label{fig:radiusplot}
\end{figure}

This work and \citet{Musoke2022} have demonstrated that disk tearing can lead to short luminosity outbursts of length $\Delta t \sim 1000-3000 r_g/c$ every $\Delta t_{tear} \sim 10,000-50,000 r_g/c$. Coincident with a tearing event the disk precesses on a timescale of $\Delta t_{prec} \sim 1000-6000 r_g/c$ suggesting tearing events might be accompanied by a sinusoidial modulation of the light curve. In parallel, the mass-weighted radius of the precessing inner disk oscillates on a timescale of $\Delta t_{osc} \sim 150-400 r_g/c$ suggesting another sinusoidal modulation of the lightcurve albeit at a higher frequency (Fig. \ref{fig:radiusplot} and \citealt{Musoke2022}). We summarize the timescales of these variability phenomena in Table \ref{table:variability} for a BHXRB ($M_{BH}=10 M_{\odot}$), a low-mass AGN ($M_{BH} = 10^6 M_{\odot}$), and a high-mass AGN ($M_{BH} = 10^9 M_{\odot}$). Please note that these timescales might vary by at least an order of magnitude depending the location of the tearing radius, which we expect will depend on e.g. the accretion rate, black hole spin, misalignment angle, and amount of large scale magnetic flux.

Table \ref{table:variability} suggests that the timescales associated with disk tearing are consistent with variety of astrophysical phenomena (see also \citealt{Nixon2014, Raj2021_AGN}). The tearing events themselves might be associated with the heartbeat modes (on $1-100$~s timescales) detected in the BHXRBs GRS 1915+105 and IGR J17091–3624 (e.g. \citealt{Belloni2000, Neilsen2011, Weng2018, Altamirano2011}) in addition to QPEs detected in various AGN (e.g. \citealt{Miniutti2019, Giustini2020, Arcodia2021, Miniutti2022}). If the tearing events in some spectral states are spaced further apart, they could potentially also explain some changing-look phenomena observed in AGN with masses of $10^7-10^9 M_{\odot}$ that occur on timescales of years to decades (e.g. \citealt{Sniegowska2020}). In all of these cases the luminosity fluctuates by a factor $10^{0}-10^{2}$ suggesting some significant flaring event in the accretion disk reminiscent of a tearing event. The precession of the disk that follows each tearing event can potentially explain low frequency QPOs detected in numerous BHXRB with $\nu \sim 10^{-1}-10^{1}$~Hz (e.g. \citealt{IngramMotta2020}). Precession is also consistent with the $44$ day LFQPO detected in the $2 \times 10^8 M_{\odot}$ AGN KIC 9650712 \citep{Smith2018} and the $3.8H$ LFQPO detected in the $10^5-10^6 M_{\odot}$ AGN GJ1231+1106 \citep{Lin2013}. In addition, the oscillation of the mass-weighted radius might be associated with HFQPOs of $\nu \sim 10^1-10^2$~Hz in BHXRBs and some of the short duration QPOs detected in various AGN. These include the $\sim 1H$ QPO in the lightcurve of the $10^6-10^7 M_{\odot}$ AGN RE J1034+396 \citep{Gierlinski2008, Czerny2016}, the $1.8H$ QPO in the $5 \times 10^6 M_{\odot}$ MRK 766 \citep{Zhang2017}, and the $2H$ QPO in the $4 \times 10^6 M_{\odot}$ AGN MS 2254.9-3712 \citep{Alston2015}.

The heated gas at the tearing radius suggests the QPO emission will be substantially hardened through Comptonization. This is consistent with frequency-resolved spectroscopy of BHXRBs which has demonstrated that the Type-C QPO waveform is dominated by hardened emission \citep{Sobolweska2006} and recent observations of AGN that suggest an increase in the soft-X-ray excess during QPEs \citep{Giustini2020}. Interestingly, our simulations suggest that tearing events are accompanied by a low-frequency sinusoidal signal caused by precession of the inner disk and a high-frequency sinusoidial signal caused by a radial epicyclic oscillation at the tearing radius. This can potentially be tested using observational data for both BHXRBs and AGN.

\begin{table}
\begin{center}
\caption{Time period between large tearing events, $\Delta t_{tear}$, disk precession period, $\Delta t_{prec}$, and period of radial epicyclic oscillations, $\Delta t_{osc}$, for various black hole masses. These timescales can be associated with various variability phenomena in BHXRBs and AGN such as QPEs, QPOs, and heartbeat oscillations (Sec.~\ref{sec:discussion}). }
\label{table:variability}
\begin{tabular}{|c|c|c|c|}
\hline
\textbf{$M_{BH}$} &  \textbf{$\Delta t_{tear}$} & \textbf{$\Delta t_{prec}$} & $\Delta t_{osc}$\\
\hline
N/A & $50,000 r_g/c$ & $6000 r_g/c$ & $400 r_g/c$ \\
\hline
$10M_{\odot}$ & $2.5s$ & $0.3s$ & $20 ms$\\
\hline
$10^6 M_{\odot}$ & $70h$ & $9h$ & $0.5h$\\
\hline
$10^9 M_{\odot}$ & $8yr$ & $1yr$ & $20d$\\
\hline
\end{tabular}
\end{center}
\end{table}

\subsection{Comparison to Semi-analytical Models}
Semi-analytical models typically divide warped accretion disks into two categories (e.g. \citealt{Papaloizou1983}) depending on the scale height of the disk relative to the $\alpha-$viscosity parameter. In the first category bending waves excited by oscillating radial and vertical pressure gradients in the warp are damped and the warp is propagated through viscous diffusion only. This happens when the scaleheight is smaller than the viscosity parameter, e.g. $h/r<\alpha$. In this regime instability criteria were derived that predict the tearing of accretion disks (e.g. \citealt{Ogilvie2000, Dogan2018, Dogan2020}). In the second category the warp is propagated through bending waves that travel at a fraction of the sound speed \citep{Papaloizou1995}. These bending waves are excited by pressure gradients in a warp. Past work has made substantial progress in describing warp propagation in the bending wave regime (e.g. \citealt{Ogilvie2013_parametric, Paardekooper2018, Deng2020,  Ogilvie2022}), and recently also addressed disk tearing in the bending wave regime using SPH simulations and linearised fluid equations \citep{Drewes2021}.

\begin{figure}
    \centering
    \includegraphics[clip,trim=0.0cm 0.0cm 0.0cm 0.0cm,width=\columnwidth]{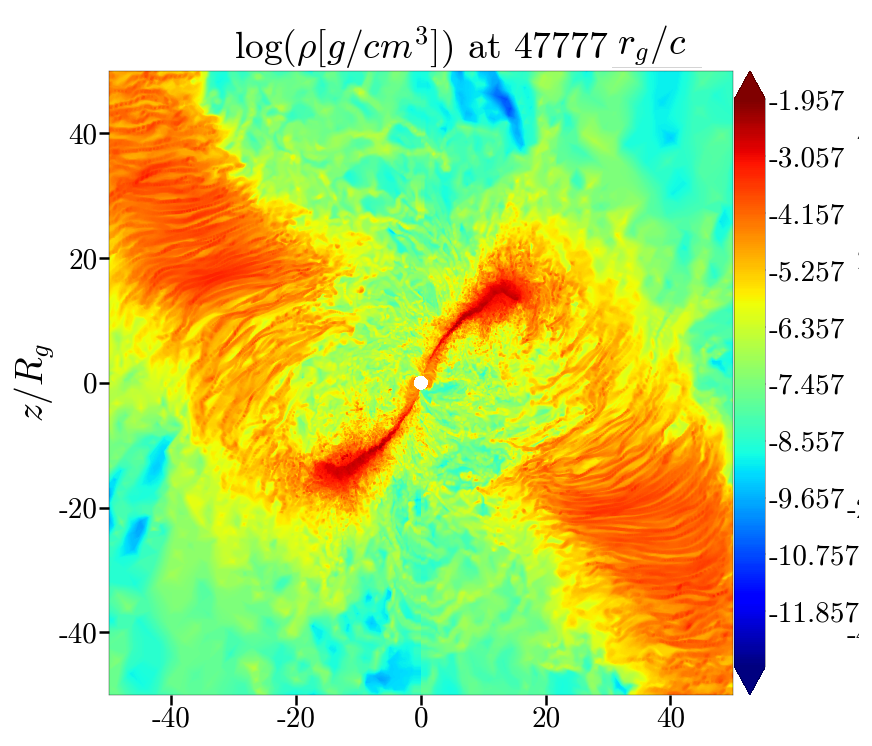}
    \vspace{-10pt}
    \caption{A transverse slice through the density along the line of nodes at $50 r_g$. This snapshot is taken beyond the `official' duration of our simulation (Sec.~\ref{sec:Physical_Setup}), just after the onset of the next tearing event. Outwards propagating wave-like structures (visible as vertical `stripes') form in the outer disk suggesting that wave-like angular momentum transport might be present. The vertical extent of the outer disk is artificially inflated because the frame is sliced at an angle with respect to the disk.}
    \label{fig:waves}
\end{figure}

Interestingly, \citet{Deng2022} predicted using the affine model of \citet{Ogilvie2018} that oscillations in the vertical pressure gradient lead to an azimuthal $m=2$ oscillation in the scale height of the disk similar to what is observed in our work. These authors also demonstrated that a parametric instability can induce turbulence and create ring-like structures in warped accretion disks. Since $h/r \sim \alpha_m$ (Fig.~\ref{fig:radplot}d) and the warp is highly non-linear it is unclear in which regime of warp propagation our disk falls (see also \citealt{Sorathia2013a, Hawley2019}). We do note though that there is some evidence for wave-like structures in the disk (Fig.~\ref{fig:waves}), which suggests the disk might be in the bending wave regime of warp propagation. Nevertheless, future work will need to isolate wave-like from diffusive (misaligned) angular momentum transport to make more direct comparison to semi-analytical models possible. 

\begin{figure}
    \centering
    \includegraphics[clip,trim=0.0cm 0.0cm 0.0cm 0.0cm,width=\columnwidth]{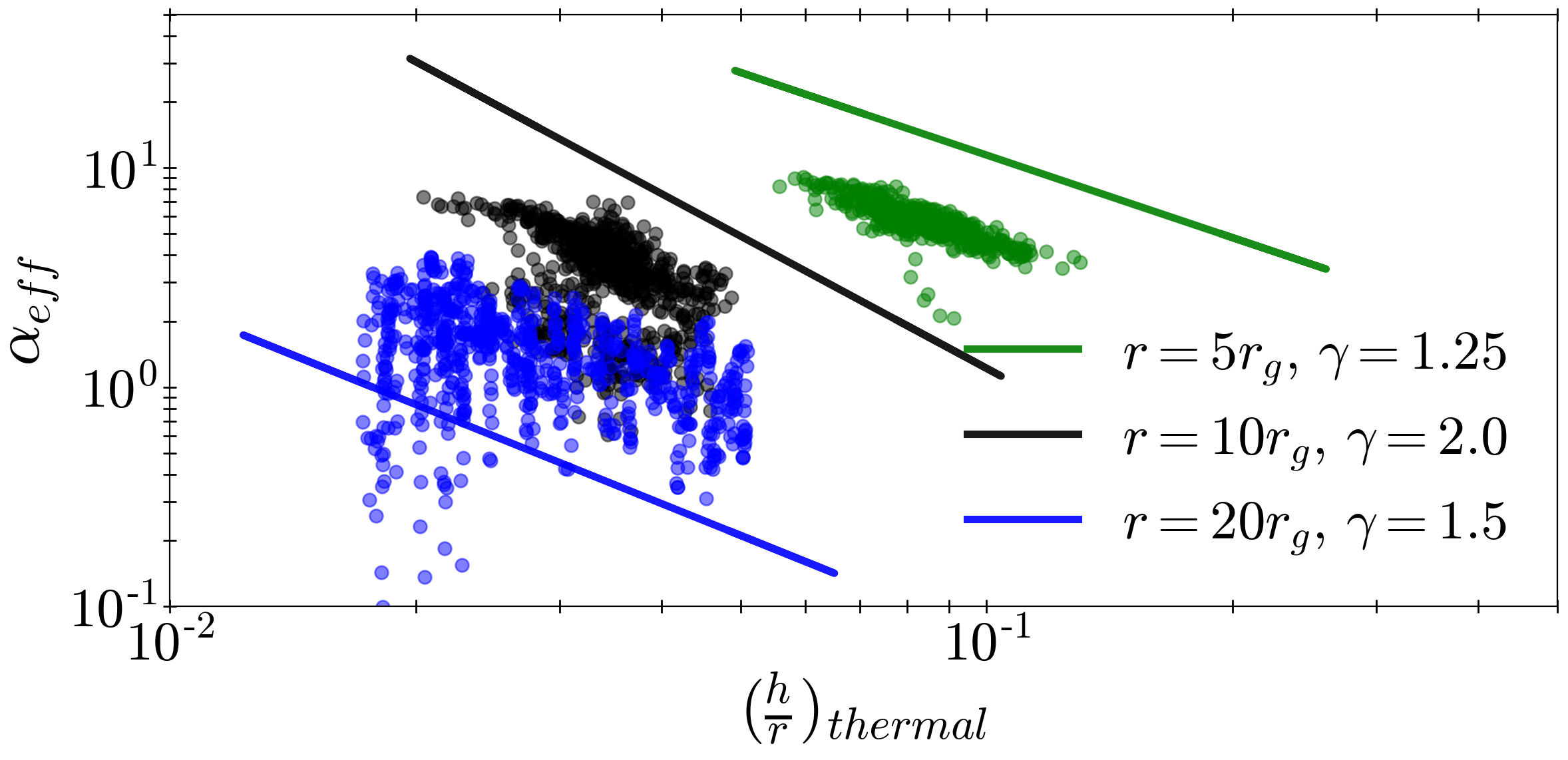}
    \vspace{-10pt}
    \caption{A scatterplot of the effective viscosity, $\alpha_{\rm eff}$, with respect to the thermal scaleheight, $\theta_{t}$ at $r=5 r_g$ in green, $r=10 r_g$ in black, and $r=20 r_g$ in blue. We show power-law fits to the data as solid lines with fitting function $\alpha_{\rm eff} \propto (h/r)^{-\gamma}$. The effective viscosity increases as the disk becomes thinner, which suggests warp-driven dissipation becomes more important for thinner disks. We argue in Section~\ref{sec:stability2} that warp-driven dissipation can stabilize the disk against runaway thermal collapse (e.g. \citealt{Shakura1976}).}
    \label{fig:alphaplot}
\end{figure}

\subsection{Thermal and Viscous Stability of Warped Disks}
\label{sec:stability2}
In Section \ref{sec:stability} we demonstrated that warped, radiation pressure supported, accretion disks remain thermally stable for much longer than similar non-warped, radiation pressure supported, accretion disks considered in previous work \citep{Liska2022}. We concluded that neither magnetic pressure nor advection of radiation internal energy can explain this thermal stability. Here we propose an alternative explanation.

For a disk to remain thermally stable the derivative of the dissipation rate, $Q_{vis}$, with respect to the temperature needs to be smaller than the derivative of the cooling rate, $Q_{rad}$, with respect to the temperature such that $\frac{Q_{vis}}{dT} < \frac{Q_{rad}}{dT}$. For a radiation pressure supported $\alpha-$disk $Q_{vis} \propto T^{r}_{\phi} H \propto \alpha p_{tot} H \propto \alpha p_{tot}^2 \propto \alpha T^8$ and $Q_{rad} \propto T^4$ \citep{ss73, meier2012}. This suggests that with a constant $\alpha$-viscosity radiation pressure supported accretion disks are thermally unstable. However, if the dissipation rate follows $Q_{vis} \propto T^{\zeta}$ with $\zeta \lesssim 4$ due to warping, the disk will remain thermally stable. 

To determine $\zeta$ in our simulation we present in Figure \ref{fig:alphaplot} a scatter plot of the effective viscosity, $\alpha_{\rm eff}$, with respect to the thermal scale height, $\theta_{t}$. We only use data after $32,000 r_g/c$ to allow the disk to adjust after seeding it with radiation and exclude data where the tilt angle is smaller than $\mathcal{T}<60^{\circ}$. We visually fit the data to find that $\alpha_{\rm eff} \propto H^{-\gamma}$ with $\gamma \sim 1.25 - 1.75$. From the relation $H \propto p_{tot} \propto T^4$  we find that $\zeta= 8 - 4 \gamma \sim 1-3$. This suggests that even though our accretion disk is radiation pressure dominated at $r=20 r_g$, warp-driven dissipation can potentially prevent the disk from becoming thermally unstable. 

Following a similar argument, one can show that the warp-driven dissipation also avoids the viscous \cite{LightmanEardley1974} instability of radiation pressure dominated disks. If a disk is subject to the viscous inability, it will disintegrate into rings of high density gas (e.g. \citealt{Mishra2016}). Even though high-density rings do form in our simulation (Figs.~\ref{fig:ST} and \ref{fig:vertical_plot}), we conclude that this cannot be a manifestation of the viscous instability since similar rings form in the non-radiative analogue to RADT65 presented by \citet{Musoke2022}. This suggests that the ring formation in RADT65 is warp-induced.

\begin{figure*}
    \centering
    \includegraphics[width=1.0\linewidth,trim=0mm 0mm 0mm 0,clip]{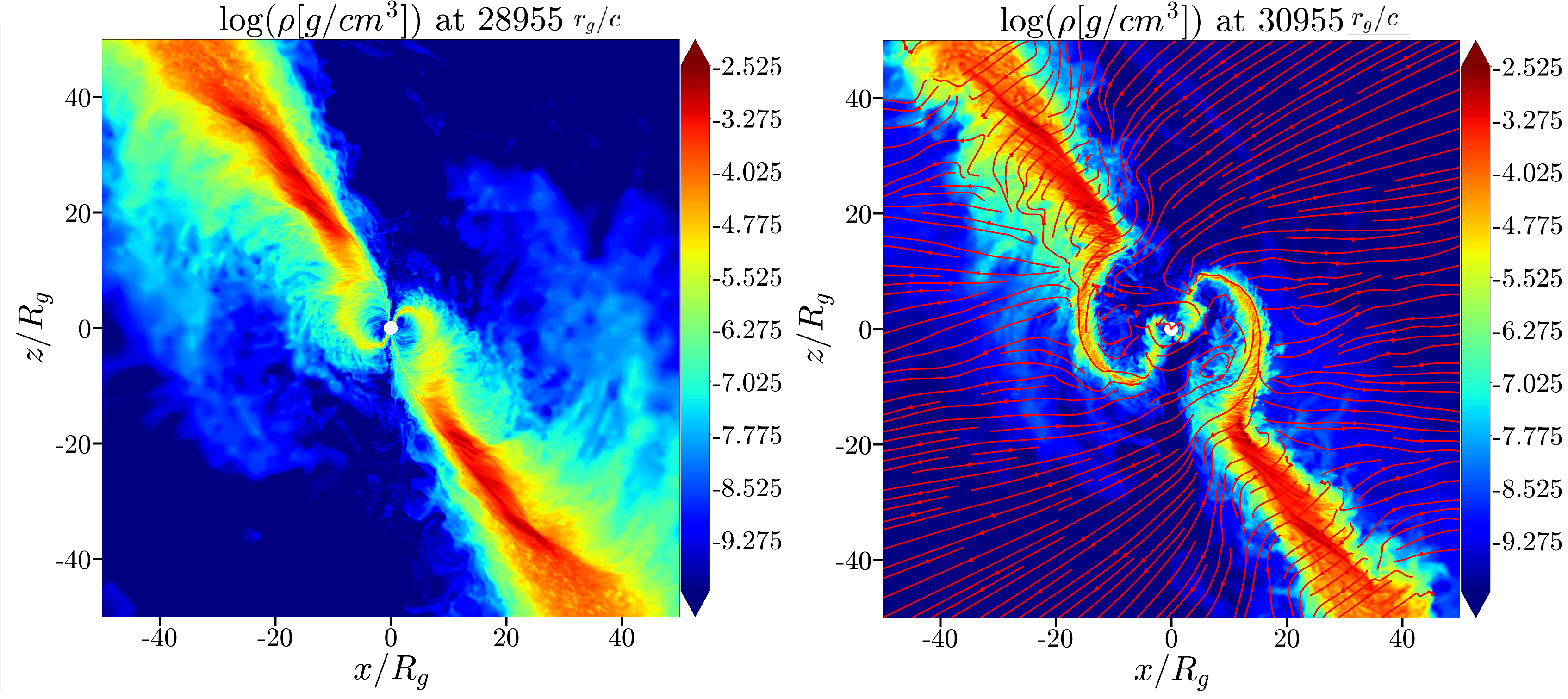}
    \caption{Radiation field warps the inner disk, as seen in a transverse slice of density through the $x-z$ plane of our initial conditions (left) and after $\Delta t=2,000 r_g/c$ (right) with radiation streamlines shown in red. Even though our M1 radiation scheme is unable to properly model multi-beam radiation (Sec.~\ref{sec:warping}), these results suggest that radiation warping might be an important effect.}
    \label{fig:rad_feedback}
\end{figure*}

\subsection{Radiation Warping}
\label{sec:warping}
When the inner disk precesses it will periodically beam radiation towards the outer disk. This can potentially warp the outer disk and induce additional dissipation (similar to \citealt{Pringle1997, Wijers1999}). In a similar context \citet{Liska2019b} has demonstrated that when a precessing jet collides with the outer accretion disk it can inject energy and angular momentum into the disk, leading to an increase of particle orbits in the outer disk. Since the energy efficiency of the radiative outflows in RADT65 is non-negligible ($|\eta_{rad}| \sim 14.7 \%$) we expect that radiation feedback can have a profound effect on the structure and dynamics of an accretion disk.

Radiative feedback is an interesting avenue of future research that is hard to address in this article due to limitations of our M1 radiation closure \citep{Levermore1984}. M1 works well in modeling radiative cooling and energy transport in optically thick regions, and in optically thin regions for a single radiation bundle. However, M1 is unable to properly model crossing radiation bundles. The reason for this is that M1 treats radiation as a  highly collisional fluid and is thus unable to model crossing radiation bundles in optically thin media. Instead the energy and momentum of the two radiation bundles will be summed together. While the M1 closure conserves the total amount of energy and momentum in the photon-fluid, it can lead to a nonphysical redistribution of energy and angular momentum. 

Despite these limitations we demonstrate in Figure \ref{fig:rad_feedback} that the structure of the inner accretion disk in RADT65 differs significantly after seeding it with radiation. In Figure \ref{fig:rad_feedback} the radiation streamlines get deflected by the inner disk presumably leading to an increase in the warp amplitude compared to the start of the simulation. In this snapshot the tearing process has not started yet, which suggests that radiation warping can be important even in the absence of disk tearing. While these preliminary results are encouraging, more advanced numerical simulations will be necessary to reliably address radiation warping.

\subsection{Role of Magnetic Fields}
\label{sec:mag_field}
In this work we have seeded the accretion disk with a relatively strong toroidal magnetic field of $\beta=\frac{p_{gas}+p_{rad}}{p_{b}} \sim 7$ (Sec.~\ref{sec:Physical_Setup}) in order to being able to resolve the poloidal MRI components and reach numerical convergence. Alternatively, we could have threaded the accretion disk with poloidal magnetic flux loops and chosen a higher plasma-$\beta$ in the initial conditions as we have done in the past (e.g. \citealt{Liska2019b}). However, this leads to the formation of powerful jets and would make these simulations more applicable to the hard-intermediate spectral state (e.g. \citealt{Liska2022}).

In an accompanying paper \citep{Kaaz2022} we demonstrate that the turbulent Maxwell stresses seeded by MRI turbulence fall short by 1-3 orders of magnitude in explaining the measured accretion rate. Instead, we demonstrate that accretion is driven by shock dissipation in the nozzle. This naively suggests that magnetic fields do not play an important role in the disk dynamics and thus that the presented results would be insensitive to the chosen plasma-$\beta$.

However, magnetic fields might still play an important role in communicating the differential Lense-Thirring torque throughout the accretion disk. This can have important consequences for the tearing process and the formation of the nozzle shock. In fact, SPH models (e.g. \citealt{Nixon2012B, Nixon2012counter, Nealon2015, Raj2021}), which do not include magnetic fields, find that a thin accretion disk tears apart into multiple deferentially precessing rings instead of a radially extended subdisk observed in our GRMHD simulations. In addition, SPH models do not find a nozzle shock. These arguments suggest that magnetic fields might play an important, though indirect role, in both the tearing process and the formation of the nozzle shock, which needs to be addressed in future work.

\subsection{Numerical Convergence}
\label{sec:convergence}
RADT65 is the largest radiative 2T GRMHD simulation to date, featuring on average over $3 \times 10^9$ cells, and requiring over $6000$ V100 GPUs on OLCF Summit. Nevertheless, to reduce the numerical cost of this work and bring it within the realm of possibilities, we needed to downscale the resolution by a factor of 2 compared to similar work presented in \citet{Liska2020} and \citet{Musoke2022} that did not take radiation into account. Despite this downscaling, the number of cells per MRI-wavelength, $Q^{i}=\lambda_{MRI}^{i}/N^{i}$, still exceeds $(Q^{\theta}, Q^{\phi}) \gtrsim (20,125)$ and the number of cells per disk scaleheight, $Z$, exceeds $Z \gtrsim 14$.

This suggests, based on more controlled convergence studies (e.g. \citealt{Shiokawa2012, Porth2019}), that both the disk and the MRI turbulence are reasonably well resolved and close to achieving convergence. In fact, the vast majority of GRMHD simulations performed to-date (e.g. \citealt{Porth2019}) have similar or even slightly inferior $Q$ and $Z$ parameters. In addition, physical quantities like the inflow speed of the gas and the evolution of the tearing radius, are very similar to the non-radiative GRMHD models in \citet{Liska2020, Musoke2022} suggesting the results presented in this article are robust and no significant qualitative changes are expected at higher resolutions.

Despite these arguments a full convergence study will be an interesting use-case for future generations of GPU clusters. Namely, doubling the resolution is currently computationally infeasible. Doing so, would $16-$fold the cost of the presented simulations to over $4$ million node hours on OLCF Summit. This would exceed the size of our entire INCITE allocation by more than an order of magnitude. On the other hand, we already know based on previous convergence studies  (e.g. \citealt{Shiokawa2012, Porth2019}) that reducing the resolution by a factor of 2 would under-resolve both the poloidal MRI components and the scaleheight of the disk, and thus fail to reach convergence.

\section{Conclusion}
\label{sec:conclusion}
In this article we have presented the first radiative two-temperature GRMHD simulation of an $65^{\circ}$ misaligned disk accreting at $\dot{M} \sim 0.35 \dot{M}_{edd}$. Similar to the idealized GRMHD models presented in \citet{Liska2019b, Liska2020, Musoke2022} the radiation pressure supported accretion disk tears apart and forms a precessing disk. During the tearing process the mass accretion rate increases by a factor $\sim 10-40$ and the luminosity increases by a factor $\sim 5$, almost exceeding the Eddington limit at the peak of the outburst. As proposed in \citet{Raj2021_AGN}, the timescales and amplitudes of these luminosity swings are roughly consistent with QPEs in AGN (e.g. \citealt{Miniutti2019, Giustini2020, Arcodia2021, Miniutti2022}) and several heartbeat modes in BHXRBs  (e.g. \citealt{Belloni2000, Neilsen2011, Altamirano2011, Weng2018}). Following a tearing event, the inner disk precesses for several periods with a frequency that is consistent with low-frequency QPOs (e.g. \citealt{IngramMotta2020}). While precessing, the inner disk also exhibits a radial epicyclic oscillation of its center of mass whose frequency is consistent with high frequency QPOs (e.g. \citealt{IngramMotta2020}). Future long-duration GRMHD simulations combined with radiative transfer calculations will need to test if disk tearing indeed produces QPO-like variability. This will require the tearing radius to remain stable over many cycles of tearing.

Warping of the disk forms two nozzle shocks directed perpendicular to the line of nodes. At the nozzle shock, the scale height of the disk decreases by a factor of $\sim 5$ and the temperature increases up to $T_e \sim 10^8-10^9 K$ and $T_i \sim 10^9-10^{10}K$. In a companion paper \citep{Kaaz2022}, we demonstrate that dissipation in the nozzle shock leads to a much shorter accretion time scale than predicted by $\alpha$-viscosity based models (e.g. \citealt{ss73}) seeded by magnetized turbulence (e.g. \citealt{Balbus1991}). This can potentially explain the timescales associated with some changing look phenomena in AGN including QPEs. In addition, we find during tearing events that gas crossing the tear heats up to $T_e \sim 10^{8} K$ and $T_i \sim 10^{10} K$. This is caused by an increase in the dissipation rate as the gas undergoes a rapid orbital plane change (see also \citealt{Nixon2014, Raj2021_AGN}). These results imply that the emission profile from warped accretion disks will deviate substantially from the \citet{Novikov1973} model of a geometrically thin accretion disk. This can, pending future radiative transfer calculations, have far-reaching consequences for spectral fitting and black hole spin measurements (e.g. \citealt{Zhang1997, Kulkarni2011}). 

While radiation pressure supported accretion disks in GRMHD simulations typically collapse (e.g. \citealt{Sadowski2016_mag, Fragile2018, Jiang2019, Mishra2020, Liska2022}) into an infinitely thin slab due to the thermal instability \citep{Shakura1976}, this did not happen in our simulation. Here we argue that the warp and associated nozzle shock stabilize the disk against thermal collapse (Sec.~\ref{sec:discussion}). These shocks differ from spiral density wave induced shocks (e.g. \citealt{Arzamasskiy2018}). In a companion paper \citep{Kaaz2022}, we demonstrate that nozzle shock dissipation drives accretion in geometrically thin warped accretion disks within at least $r \lesssim 50 r_g$. This upends the standard paradigm that invokes magneto-rotational instability induced turbulence for the dissipation of orbital kinetic energy (e.g. \citealt{Balbus1991, Balbus1998}). Future work will need to address the range of spectral states, accretion rates, and misalignment angles where nozzle shock dissipation becomes important. 

Past work almost exclusively invoked magnetic reconnection in coronae to explain high energy (non-)thermal emission from accretion disks (e.g. \citealt{Sironi_2014, Beloborodov2017, RipperdaLiskaEtAl2021}). However, this work suggests that heated plasma formed in the nozzle and at the tearing radius can potentially be another source of high energy emission. This is especially appealing to explain non-thermal emission in the high-soft and soft-intermediate states of BHXRBs. Those states do not launch any large-scale jets and are thus unlikely to be saturated by magnetic flux (e.g. \citealt{Mckinney2012}), which might be a key ingredient to form a corona where the prime source of dissipation comes from magnetic reconnection (e.g. \citealt{Liska2022, RipperdaLiskaEtAl2021}). Test-particle methods (e.g. \citealt{Bacchini2019}) informed by first principles particle-in-cell simulations are an interesting avenue for future research to quantify the contribution of magnetic reconnection and warp-driven shocks to the acceleration of non-thermal particles and corona-like emission. 

\section{Acknowledgements}
We thank Michiel van der Klis, Adam Ingram, Sera Markoff and Ramesh Narayan for insightful discussions.  An award of computer time was provided by the Innovative and Novel Computational Impact on Theory and Experiment (INCITE), OLCF Director's Discretionary Allocation, and ASCR Leadership Computing Challenge (ALCC) programs under award PHY129. This research used resources of the Oak Ridge Leadership Computing Facility, which is a DOE Office of Science User Facility supported under Contract DE-AC05-00OR22725. This research used resources of the National Energy Research
Scientific Computing Center, a DOE Office of Science User Facility
supported by the Office of Science of the U.S. Department of Energy
under Contract No. DE-AC02-05CH11231 using NERSC award
ALCC-ERCAP0022634. We acknowledge PRACE for awarding us access to JUWELS Booster at GCS@JSC, Germany. ML was supported by the John Harvard Distinguished Science Fellowship, GM is supported by a Netherlands Research School for Astronomy (NOVA), Virtual Institute of Accretion (VIA) postdoctoral fellowship, NK by an NSF Graduate Research Fellowship, and AT by the National Science Foundation grants AST-2206471, AST-2009884,AST-2107839, AST-1815304, OAC-2031997, and AST-1911080.

\newpage

\bibliography{new.bib,extra.bib}{}
\bibliographystyle{aasjournal}



\end{document}